\documentclass[aps,prd,twocolumn,groupedaddress,nofootinbib]{revtex4-1}
\usepackage[utf8]{inputenc}
\usepackage[english]{babel}
\usepackage[T1]{fontenc}
\usepackage{amsmath,amsfonts,amsthm,bm}
\usepackage{amsfonts}
\usepackage{hyperref}
\usepackage{amssymb}
\usepackage{graphicx}
\usepackage{tikz-feynman}
\usepackage{subfig}
\usepackage{physics}
\usepackage{array}
\usepackage[normalem]{ulem}

\newcommand{\mpl}{M_\mathrm{P}}

\begin{document}

% Use the \preprint command to place your local institutional report
% number in the upper righthand corner of the title page in preprint mode.
% Multiple \preprint commands are allowed.
% Use the 'preprintnumbers' class option to override journal defaults
% to display numbers if necessary
%\preprint{}

%Title of paper
\title{Effective two-body approach to the hierarchical three-body problem}

% repeat the \author .. \affiliation  etc. as needed
% \email, \thanks, \homepage, \altaffiliation all apply to the current
% author. Explanatory text should go in the []'s, actual e-mail
% address or url should go in the {}'s for \email and \homepage.
% Please use the appropriate macro foreach each type of information

% \affiliation command applies to all authors since the last
% \affiliation command. The \affiliation command should follow the
% other information
% \affiliation can be followed by \email, \homepage, \thanks as well.
\author{Adrien Kuntz}
\email[]{adrien.kuntz@sns.it}
\author{Francesco Serra}
\email[]{francesco.serra@sns.it}
\author{Enrico Trincherini}
\email[]{enrico.trincherini@sns.it}
%\homepage[]{Your web page}
%\thanks{}
%\altaffiliation{}
\affiliation{Scuola Normale Superiore, Piazza dei Cavalieri 7, 56126, Pisa, Italy}
\affiliation{INFN Sezione di Pisa, Largo Pontecorvo 3, 56127 Pisa}

%Collaboration name if desired (requires use of superscriptaddress
%option in \documentclass). \noaffiliation is required (may also be
%used with the \author command).
%\collaboration can be followed by \email, \homepage, \thanks as well.
%\collaboration{}
%\noaffiliation

\date{\today}

\begin{abstract}
The motion of three bodies can be solved perturbatively when a tightly bound inner binary is orbited by a distant perturber, giving rise for example to the well-known Kozai-Lidov oscillations. 
 We propose to study the relativistic hierarchical three-body orbits by adapting the Effective Field Theory techniques used in the two-body problem. This allows us to conveniently treat the inner binary as an effective point-particle, thus reducing the complexity of the three-body problem to a simpler spinning two-body motion. We present in details the mapping between the inner binary osculating elements and the resulting spin of the effective point-particle. Our study builds towards a derivation of three-body analytic waveforms.

\end{abstract}

% insert suggested PACS numbers in braces on next line
%\pacs{}
% insert suggested keywords - APS authors don't need to do this
%\keywords{}

%\maketitle must follow title, authors, abstract, \pacs, and \keywords
\maketitle

\section{Introduction}

Triple system in nature often come in hierarchical configurations~\cite{Naoz_2016}. In this kind of setting, a close (or "inner") binary $m_1$-$m_2$ is orbited by a distant perturber $m_3$ (the "outer" object). Studies of systems of this kind include satellites and asteroids in the Solar system~\cite{2015aste.book..355M, Nesvorn__2011}, triple stars~\cite{2014AJ....147...87T, Stephan_2016, Naoz_2014, 1997AstL...23..727T}, exoplanets~\cite{2014ApJ...785..126K, Veras_2010, Wu_2007, Ngo_2015} or triple black holes and neutron stars systems~\cite{Randall:2018nud, Seto:2013wwa, Antonini_2016, 10.1093/mnras/stt617, Deane:2014jqa}.
 The first study of hierarchical triples dates back to Lidov and Kozai~\cite{1962AJ.....67..591K, LIDOV1962719}. They discovered that the triple system evolves also on a characteristic timescale much longer than the period of the two orbits, which is now referred to as the Kozai-Lidov (KL) timescale:
\begin{equation}
T_\mathrm{KL} \simeq \frac{T_3}{T} T_3 \; ,
\end{equation}
where $T_3$ (resp. $T$) is the period of the outer (resp. inner) orbit. On this long timescale, the inner system can experience eccentricity and inclination oscillations, if it starts from a configuration with large relative inclination. Thus, hierarchical systems of black holes can feature dramatically reduced merger timescales~\cite{2018ApJ...856..140H}. Since these systems are quite common in dense stellar environments~\cite{2016ApJ...824L..12O, 2020ApJ...903...67M}, this makes them especially relevant to gravitational wave astronomy.

 The conventional treatment of the hierarchical problem proceeds by expanding the Hamiltonian in the ratio of semimajor axes, which we denote by $\varepsilon$:
\begin{equation}
\varepsilon = \frac{a}{a_3} \; ,
\end{equation}
where $a$ an $a_3$ are the semimajor axes of the inner and outer orbit respectively.
Then, one can average the expanded Hamiltonian on both orbital timescales to obtain a set of long-timescale evolution equations of orbital quantities, a procedure known as adiabatic or secular approximation. The set of equations thus obtained is commonly referred to as the Lagrange planetary equations. The quadrupole term of this averaged expansion gives rise to the KL oscillations.
The next level of approximation in $ a/a_3 $, namely the octupole, leads to even richer possibilities like orbital flips, extreme eccentricities and chaotic evolution~\cite{2011Natur.473..187N, 10.1093/mnras/stt302, Ford_2000}.

On the other hand, General Relativity (GR) brings corrections to the motion proportional to the relative velocity between the bodies $v$, which can be regarded as an expansion parameter (we will use units in which $c=1$). The interplay between two-body GR effects and Kozai-Lidov oscillations has been studied by numerous authors, e.g.~\cite{Blaes_2002, 10.1111/j.1365-2966.2007.11694.x, 10.1093/mnras/stu039, Biscani:2013zva}. Generically, the inner binary precession tends to suppress the eccentricity oscillations if the GR timescale is much shorter than the KL one~\cite{Ford_2000}. In other parts of the phase space, though, post-Newtonian corrections combined with the three-body ones can excite eccentricities~\cite{Naoz_2013}.

However, much less is known concerning the corrections brought by genuine three-body relativistic effects. These terms are essential to derive a waveform of a hierarchical three-body system (they can give rise to the so-called "tidal resonances"~\cite{Bonga_2019}) or to obtain the correct time-evolution of the system in some parts of the parameter space~\cite{2019ApJ...883L...7L}. One can approach the problem with a numerical relativistic three-body solver~\cite{PhysRevD.83.084013, PhysRevD.84.104038, 10.1093/mnras/stw1590,Lousto:2007ji,Gupta:2019unn}, however this method is usually time-consuming and inadequate for the derivation of an analytic inspiral waveform model to be used in matched filter analysis~\cite{Owen_1999}.  Another valid approach is to further assume a hierarchy of masses $m_3 \gg m_1, m_2$ so that the system can be studied with black hole perturbation theory~\cite{2019MNRAS.488.5665W,Torres_Orjuela_2019,Meiron_2017,Yu_2021, cardoso2021gravitational,Torres_Orjuela_2020,Yang_2019,Han_2019,Bonga_2019}. However, in this article we will rather focus on the similar-mass case $m_1 \sim m_2 \sim m_3$.

Most previous studies on the fully relativistic hierarchical three-body problem use a combination of the post-Newtonian formalism~\cite{Blanchet:2013haa} and the quadrupole expansion at the level of the equations of motion~\cite{Fang_2020, Fang_2019, PhysRevD.89.044043, PhysRevLett.120.191101, 2019ApJ...883L...7L, Lim:2020cvm}. Some other authors use instead a Hamiltonian formalism~\cite{Naoz_2013,Migaszewski:2008tp}. However, their studies do not take into account the so-called {\it adiabatic corrections} computed by Will~\cite{PhysRevD.89.044043,PhysRevLett.120.191101} and Lim and Rodriguez~\cite{Lim:2020cvm}. The computations needed to get the "cross-terms" representing the interaction of the multipole expansion with relativistic corrections can be quite cumbersome, and it may be difficult to gather physical intuition from the results of the calculations.

In this paper, instead, we begin exploring three-body systems following a new approach, based on a number of powerful Effective Field Theories (EFT) techniques that have been developed for the relativistic two-body problem in recent years. The framework we build on goes under the name of Non-Relativistic General Relativity (NRGR)~\cite{goldberger_effective_2006, porto_effective_2016}, and its generalization to include the spin of the constituents~\cite{Porto_2006, Levi:2015msa, Levi:2018nxp}, which leads to the computations of new post-Newtonian terms related to spin~\cite{Porto_2006_2, Porto_2008, Porto:2008tb, Porto:2008jj, Porto_2010, Porto_2011, Porto_2012, cho2021gravitational, Levi:2014gsa,Levi:2010zu,Levi:2015ixa,Levi:2020kvb}. In the EFT language, the gravitational multipole expansion is implemented at the level of the action~\cite{ross_multipole_2012}, using symmetries to restrict the form of the allowed terms~\cite{goldberger_gravitational_2010}. The multipole expansion derived in this way has then been used to compute the gravitational dissipative dynamics in the GR two-body problem.

In the following, we will apply similar ideas to the hierarchical three-body problem. Such a system is particularly suited to an EFT description for two reasons. First, its dynamics is characterized by two small dimensionless ratios of scales,  $v$ and $\varepsilon$. The EFT power counting rules allow to estimate easily the sizes of different contributions, thus dictating to what order in perturbation they have to be computed, at a given experimental accuracy. Ratios of different scales can be kept to different orders, depending on their numerical values. In the second place, symmetries are manifest at the level of the EFT Lagrangian and they restrict the form of the allowed terms. As we will see, this considerably simplifies the form of the cross-terms compared to the existing literature and it allows to gather some physical intuition about the effects of relativistic multipole corrections to the dynamics.   

The very nature of the Effective Field Theory framework requires to first identify the hierarchy of well-separated length scales involved in a system and remove (integrate out) each of them, one at a time, starting form the smallest. In this way a tower of EFTs is obtained, that eventually leads to the infrared (long-distance) description of the problem one is interested in. 
%This also means that it will be possible to adapt the already large bank of two-body spinning templates in order to study three-body waveforms.

Thus, we will first focus on the inner binary and integrate out the gravitational field in the presence of an external perturbation, which will be ultimately generated by the third body. The resulting theory will match onto a composite particle, endowed with spin and multipole moments, coupled to gravity. Such a treatment will be valid away from resonances~\footnote{If the perturbation was in resonance with the modes of the inner orbital motion, then it would be much more difficult to integrate out these modes and to describe the inner binary as an effective point particle. The same obstacle is encountered in the double averaging procedure, see for instance Appendix A2 of Ref.~\cite{10.1093/mnras/stt302}.}, and as long as the ratio of semimajor axes $\epsilon$ remains small at all times. This procedure means replacing a three-body problem with a simpler two-body one, where one of the two point-particles is the inner binary.
% Repeating it another time, we will obtain a description in terms of two interacting composite particles, the inner and the outer orbit. {\color{blue}We not not really treat the outer binary as a point-particle... I would remove this last sentence, or maybe replace it by : It is then straightforward to use the standard tools of the two-body problem with spinning point-particles to compute the evolution of a three-body system.}

Here we describe briefly the distinct steps to guide the reader in the rest of the paper. 
\begin{enumerate}
\item We will start from a system of three worldlines minimally coupled to gravity, where we have already integrated out the modes whose wavelengths are comparable with the size of the bodies. From this starting point, we will integrate out the off-shell modes that contribute to the gravitational potential of the inner binary, having momenta $ k^{\mu}\sim(v/a,1/a) $. Thus we will obtain an action describing the gravitational interaction of the two inner bodies in the presence of an external gravitational field. This will be done in Section~\ref{sec:Leff}. 
\item Then we will first expand the Lagrangian in multipoles and, after that, since we are interested in long-time scale evolution, we integrate out the point-particle orbital modes with frequencies $ \omega > v/a $. In practice this will be done by averaging over the period of the inner orbit. Doing so, we will obtain the action of a composite particle, whose spin is simply the orbital angular momentum of the inner binary, coupled to an external gravitational field. This step will be carried out in Section~\ref{sec:multipolar}. Although the final result may seem straightforward from an EFT perspective (gauge invariance fixes all the terms in the action to dipolar order without any free parameter, so that the matching might seem superfluous), the computation will allow us to find the exact relation between the center-of-mass choice and the so-called "spin supplementary condition" (SSC), which is a particular gauge choice for the spin tensor.
\item Similarly to the first step, we will then consider the two worldlines, one for the third body and the other for the composite spinning particle representing the inner binary.  We introduce the "effective two-body" EFT and show explicitly its power-counting rules in both expansion parameters $v$ and $\varepsilon$. Integrating out the off-shell modes with momenta $ k^{\mu}\sim(V/a_3,1/a_3) $, we will obtain an action describing the gravitational interaction between the inner binary and the third body.
\item Finally, we will integrate out the remaining  point-particle orbital modes with frequencies $ \omega > V/a_3 $, doing an average over the period of the outer orbit. In this way we will get to a Lagrangian representing the dynamics of the 3-body system as an interaction between the composite particle representing the inner binary and the outer body. These last two steps will be carried out in Section~\ref{sec:integrate_outer}.
\end{enumerate}

%
%Let us now be more precise about the content of this article. 
%%In Section~\ref{sec:LPE}, we will introduce the basic building block of the hierarchical three-body problem in its usual treatment, namely the Lagrange Planetary Equations (LPE). We will see that this set of equations can be recast, in the adiabatic approximation, as a spin kinetic term for the angular momentum of the binary.
%In Section~\ref{sec:Leff}, we will carry out the step 1 of our procedure. We will also set the stage for later computations, commenting on the relativistic definition of the center-of-mass and introducing the osculating elements that describe the perturbed motion of the binary.
%In Section~\ref{sec:multipolar} we will cover the step 2 and match the inner binary to an effective point-particle, whose spin is the orbital angular momentum of the system. This will be done by first taking a multipolar expansion of the interaction between the inner binary and the external gravitational field and then averaging over the period of the inner orbit. In Section~\ref{sec:integrate_outer} we will carry out the last two steps of our plan, discussing the "effective two-body" theory and deriving the coupling between the two orbits.

Besides these points, in Section~\ref{sec:Leff} we will also comment on the relativistic definition of the center of mass and introduce the osculating elements that describe the perturbed motion of the binary. We elaborate on the relation between the spin kinetic term and the Lagrange planetary equations in Appendix~\ref{app:LPE}, while in Appendix~\ref{app:spin} we provide details about the specific spin supplementary condition used in this article.

For the moment being, we will carry out our computations up to dipole and 1PN order. Already at this stage, we will highlight a number of conceptual clarifications arising from the EFT treatment. However, several interesting new terms also arise at quadrupolar order, related to the corrections to adiabatic approximation~\cite{PhysRevD.89.044043, Lim:2020cvm}. While we will briefly comment on the allowed form of these terms (restricted by symmetries) in this paper, we will defer a complete study of them to further work.

We will use the mostly positive metric signature. Planck's mass is defined by $\mpl^2 = 1/(8\pi G_N)$. Given the numerous different symbols used in this article, we provide here a dictionary of our notation:
\begin{itemize}
\item $\bm y_1$, $\bm v_1$, $\bm y_2$, $\bm v_2$: positions and velocities of the two constituents of the inner orbit, of masses $m_1$ and $m_2$;
\item $\bm y_3$, $\bm v_3$: position and velocity of the external perturber, of mass $m_3$;
\item $\bm Y_\mathrm{CM}$, $\bm V_\mathrm{CM}$: position and velocity of the center-of-mass of the inner binary, defined in Section~\ref{sec:CM}
\item $\bm r = \bm y_1 - \bm y_2$, $r = \vert \mathbf{r} \vert$, $\mathbf{n} = \mathbf{r}/r$ $\bm v = \bm v_1 - \bm v_2$, $\bm R = \bm Y_\mathrm{CM} - y_3$, $R = \vert \mathbf{R} \vert$, $\mathbf{N} = \mathbf{R}/R$, $\bm V = \bm V_\mathrm{CM} - v_3$;
\item $m=m_1 +m_2$ is the mass of the inner binary, $M=m_1 +m_2+m_3$ is the total mass of the system, $\mu = m_1m_2/m$ is the reduced mass of the inner and $\nu = \mu/m$ its symmetric mass ratio. Similarly, $\mu_3 = m_3 m/M$ and $\nu_3 = \mu_3/M$ are the reduced mass and symmetric mass ratio of the outer;
\item $a$ [$a_3$]: semimajor axis of the inner [outer] orbit;
\item $e$ [$e_3$]: eccentricity of the inner [outer] orbit;
\item $\bm{\hat \alpha}$, $\bm{\hat \beta}$, $\bm{\hat \gamma}$ [$\bm{ \hat \alpha}_3$, $\bm{ \hat \beta}_3$, $\bm{ \hat \gamma}_3$]: orthonormal basis of vectors characterizing the inner [outer] orbit, aligned respectively along the semimajor axis (pointing towards the pericenter), the semiminor axis, and the angular momentum;
\item $\Omega$, $\omega$, $\iota$: angles characterizing the orientation of the inner orbit (the "orbital elements"), defined by $\hat{ \bm{\alpha}} = R_z(\Omega) R_x(\iota) R_z(\omega) \hat{ \mathbf{u}}_x$ where the $R_{x_i}$'s are rotation matrices along the given axis $ x_i $;
\item $u$ [$\eta$]: mean [eccentric] anomaly of the inner orbit;
\item $L$, $G$, $H$: conjugate momenta to $u$, $\omega$ and $\Omega$ respectively, defined in Eq.\eqref{eq:Lagrangian_planetary_angular};
%\item $n=\sqrt{G_N m / a^3}$: frequency of the inner orbit;
\item $\bm{J} = \mu \sqrt{G_N ma(1-e^2)} \hat{\bm{\gamma}}$ [$\bm{J}_3 = \mu_3 \sqrt{G_N M a_3(1-e_3^2)} \hat{\bm{\gamma}}_3$]: angular momentum vector of the inner [outer] orbit;
%\item $\bm{J} = \mu \sqrt{G_N ma(1-e^2)} \bm{\gamma}$: angular momentum vector of the inner orbit.
\item $ \mathcal{E} = m - G_N m \mu/(2a) $: Total (mass and Newtonian) energy of the inner binary;
\end{itemize}

\begin{figure}
\includegraphics[width=0.7\columnwidth]{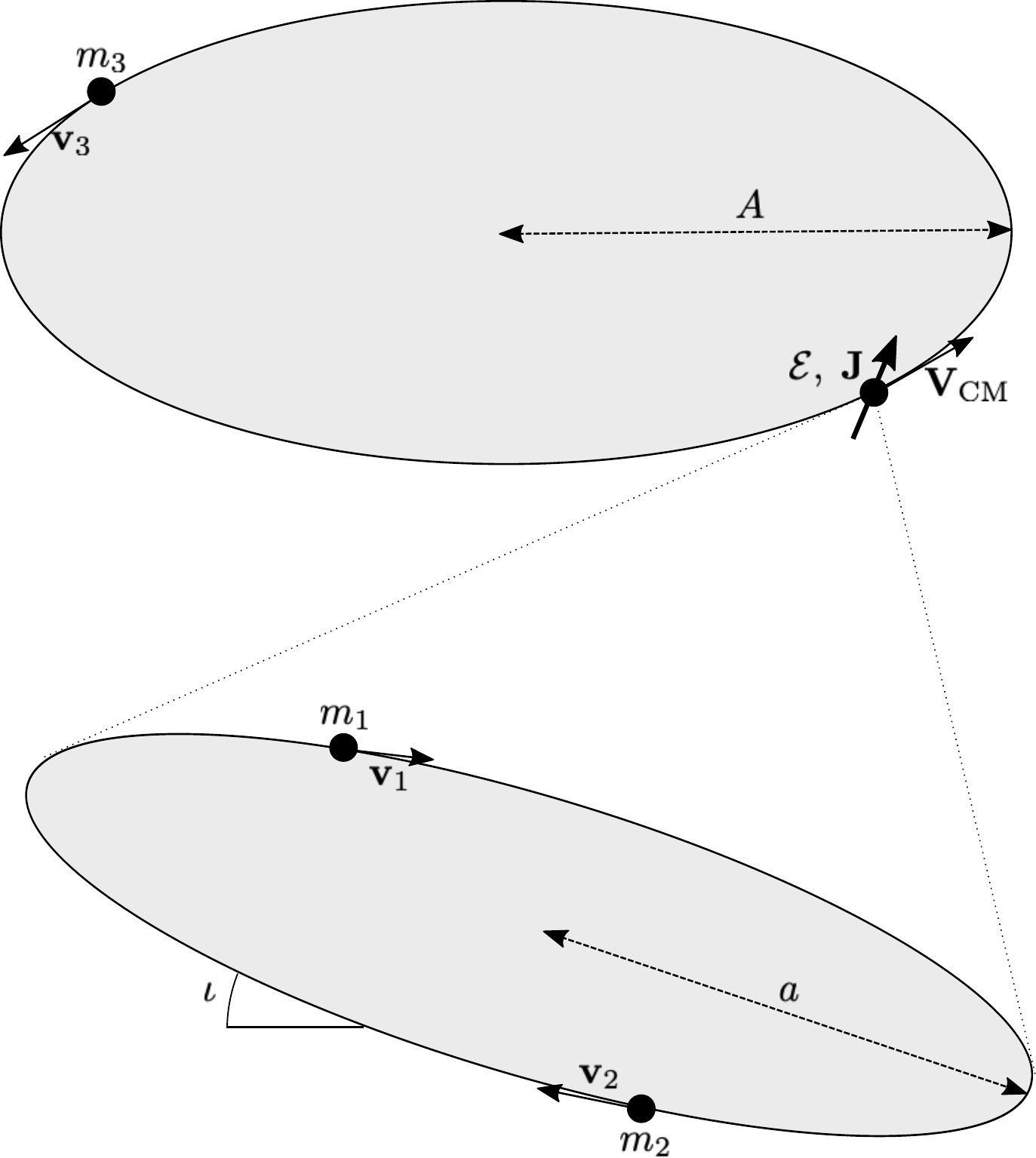}
\caption{Structure of the "effective two-body" description. The inner binary is replaced by an effective point-particle whose mass and spin are respectively the relativistic binding energy and the angular momentum of the binary system.}
\label{fig:illustration}
\end{figure}

\section{A binary system in an external field} \label{sec:Leff}

In this Section, we will obtain the effective Lagrangian at the first post-Newtonian order for the inner two-body system using the background field method, which amounts to integrate out the metric fluctuations in the presence of an arbitrary external field. Up to dipole order, we will then explicitly match this Lagrangian to the one of a spinning point-particle coupled to gravity. This spin coupling induces the dominant non-trivial post-Newtonian evolution of the inner binary parameters in the hierarchical three-body problem.

Before integrating out the gravitational field, let us introduce a convenient notation.  We will write the Lagrangian of the binary as
\begin{equation} \label{eq:def_L0}
\mathcal{L} =  \frac{1}{2} \mu v^2 + \frac{G_N \mu m}{r} + \mathcal{L}_1 \equiv \mathcal{L}_0 + \mathcal{L}_1 \; , 
\end{equation}
where $\mu = m_1 m_2/(m_1 + m_2)$ is the reduced mass, $\mathbf{r} = \mathbf{y}_1 - \mathbf{y}_2$, $\mathbf{v} = \mathbf{v}_1 - \mathbf{v}_2$ and $\mathcal{L}_1$ is called the \textit{perturbing function}. For instance, considering only the Newtonian-order perturbation due to the additional Newtonian potential $\Phi$ from the third body, the perturbing function reads 
\begin{equation} \label{eq:perturbingFunctionThirdBody}
\mathcal{L}_1 = \frac{1}{2} m V_\mathrm{CM}^2 - m_1 \Phi(t, \mathbf{y}_1) - m_2 \Phi(t, \mathbf{y}_2) \; ,
\end{equation}
where $\mathbf{V}_\mathrm{CM}$ is the (Newtonian) center-of-mass velocity. The aim of this Section is to compute the 1PN terms in the perturbing function.

%\subsection{Two-body Lagrangian in an external field}
\subsection{The Lagrangian up to 1PN order}

%\paragraph{Kaluza-Klein decomposition}

In order to make the computations as simple as possible, we will use the Kaluza-Klein decomposition space+time of the metric presented in~\cite{Kol:2007bc, Kol:2010si}, since in the NR regime the time dimension can be considered as compact in comparison to the spatial dimensions. The full metric is decomposed into a scalar $\phi$, a spatial vector $A_i$ and a spatial metric $\gamma_{ij}$ in the following way:
\begin{equation}
\mathrm{d}s^2 = - e^{2 \phi} \left(\mathrm{d}t - A_i \mathrm{d}x^i \right)^2 + e^{-2 \phi} \gamma_{ij} \mathrm{d}x^i \mathrm{d}x^j \; .
\end{equation}
We take the field action to be the standard Einstein-Hilbert term with a harmonic gauge-fixing term~\cite{goldberger_effective_2006},
\begin{equation}
S = \frac{\mpl^2}{2} \int \mathrm{d}^4 x \sqrt{-g} \; R - \frac{\mpl^2}{4} \int \mathrm{d}^4 x \sqrt{-g} \; g_{\mu \nu} \Gamma^\mu \Gamma^\nu \; ,
\end{equation}
where $\Gamma^\mu$ is the harmonic gauge condition,
\begin{equation}
\Gamma^\mu = \Gamma^\mu_{\nu \rho} g^{\nu \rho} \; .
\end{equation}
In the non-relativistic limit and in the conservative sector of the dynamics, temporal derivatives are treated as an interaction term.
Up to 1PN order we will only need the part of the action defining the $\phi$ and $A_i$ propagators, so that the action simplifies to
\begin{equation}
S = \frac{\mpl^2}{2} \int \mathrm{d}^4 x \left[ 2 (\partial_\mu \phi)^2 - \frac{1}{2} (\partial_i A_j)^2   \right] \; .
\end{equation}
Consequently, the propagators of the (Fourier-space) fields are given in the non-relativistic regime by
\begin{align}
\left\langle T \phi(\mathbf{k}, t_1) \phi(\mathbf{q},t_2)  \right\rangle &= - \frac{i}{2 \mathbf k^2 \mpl^2} \delta^3(\mathbf{k}+\mathbf{q}) \delta(t_1-t_2) \; , \\
\left\langle T A_i(\mathbf{k},t_1) A_j(\mathbf{q},t_2)  \right\rangle &=  \frac{2i}{ \mathbf k^2 \mpl^2} \delta_{ij} \delta^3(\mathbf{k}+\mathbf{q}) \delta(t_1-t_2) \; ,
\end{align}
and there is an additional scalar temporal vertex whose expression is $- \mpl^2  \int \mathrm{d}^4 x \;   \dot \phi^2$. 

To the Einstein-Hilbert term we add two point-particles $A=1,2$ whose action is
\begin{align}
\begin{split}
S_\mathrm{pp, A} &=  -m_A \int \mathrm{d}t \sqrt{-g_{\mu \nu} v_A^\mu v_A^\nu}\\
& = -m_A \int \mathrm{d}t \; e^\phi \sqrt{(1-\mathbf{A} \cdot \mathbf{v}_A)^2 - e^{- 4 \phi} v_A^2} \; ,
\end{split}
\end{align}
where $v_A^\mu = (1, \mathbf{v}_A)$ is the coordinate velocity of the point-particle. We have set $\gamma_{ij} = \delta_{ij}$ since the fluctuations of $\gamma_{ij}$ contribute only starting from 2PN order~\cite{Kol:2007bc}.
We expand the point-particle action for weak-field values. At 1PN order, the only vertices contributing are:
\begin{align}\label{eq:pp}
\begin{split}
S_\mathrm{pp, A} = - m_A \int \mathrm{d} t &\left( 1 - \frac{v_A^2}{2} - \frac{v_A^4}{8} - \mathbf{A} \cdot \mathbf{v}_A \right. \\
 &+ \left. \phi \left(1 + \frac{3}{2}v_A^2 \right) + \frac{\phi^2}{2} \right) \; .
 \end{split}
\end{align}

We now use the background field method by splitting the fields according to  $\phi = \bar \phi + \tilde \phi$, $A_i = \bar A_i + \tilde A_i$. The tilde quantities correspond to an external arbitrary field (later on, we will relate this field to the one generated by the third point-particle), while we integrate out the barred quantities corresponding to gravitons exchanges between the two bodies.
The part of the Lagrangian which does not depend on $\tilde \phi$ and  $\tilde A_i$ is the so-called EIH Lagrangian~\cite{einstein_gravitational_1938}. Since it has already been computed in this framework by several references~\cite{goldberger_effective_2006, Kol:2007bc, Kol:2010si}, we will simply give its expression without explicitly computing the relevant Feynman diagrams:
\begin{align}\label{eq:LEIH}
\begin{split}
\mathcal{L}_\mathrm{EIH} &=  \frac{1}{2} m_1 v_1^2 + \frac{1}{2} m_2 v_2^2 + \frac{G_N m_1 m_2}{r} \\ &+ \frac{1}{8} m_1 v_1^4 + \frac{1}{8} m_2 v_2^4 + \frac{G_N m_1 m_2}{2 r} \bigg[ 3 v_1^2 + 3 v_2^2  \\ &- 7 \mathbf{v}_1 \cdot \mathbf{v}_2 - \mathbf{v}_1 \cdot \mathbf{n} \; \mathbf{v}_2 \cdot \mathbf{n} - \frac{G_N m}{r} \bigg] \; ,
\end{split}
\end{align}
where $\mathbf{r} = \mathbf{y}_1 - \mathbf{y}_2$, $r = \vert \mathbf{r} \vert$ and $\mathbf{n} = \mathbf{r}/r$.

%\mathcal{L}_0 &= L_\mathrm{Newt} + L_\mathrm{EIH} \; , \\
%\mathcal{L}_\mathrm{Newt} &= \frac{1}{2} m_1 v_1^2 + \frac{1}{2} m_2 v_2^2 + \frac{G_N m_1 m_2}{r} \; , \\

\begin{figure}
	\centering
%	\subfloat[]{
%		\begin{tikzpicture}
%			\begin{feynman}
%				\vertex (i1);
%				\vertex [right=of i1] (a);
%				\vertex [right=of a] (f1);
%				\vertex [above=of f1] (s);
%				\vertex [below=of i1] (i2);
%				\vertex [below=of a] (b);
%				\vertex [below=of f1] (f2);
%				
%				\diagram*{
%				i1 -- [fermion] (a) -- [fermion] (f1),
%				(a) -- [gluon] (s),
%				(i2) -- [fermion] (b) -- [fermion] (f2),
%				(a) -- [scalar] (b)
%				};
%			\end{feynman}
%		
%		\end{tikzpicture}
%	}\hspace{1em}

\includegraphics[width=0.4\columnwidth]{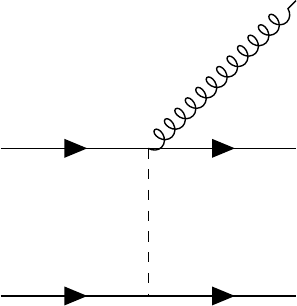}
	
\caption{Feynman diagram contributing to the emission of one scalar, at  order $v^2$.}
\label{fig:feynDiagr_phitilde}
\end{figure}

Next, including background fields, we can compute the perturbing function $\mathcal{L}_1$ defined in Eq.~\eqref{eq:def_L0}, integrating out $\bar \phi$ and $\bar A_i$. At 1PN order the result is given by:
\begin{align}\label{eq:L1PN_unexpanded}
\begin{split}
\mathcal{L}_1 &\equiv \mathcal{L} - \mathcal{L}_0 = \mathcal{L}_\mathrm{EIH} - \mathcal{L}_0 - m_1 \tilde \phi(\mathbf{y}_1) \left( 1 + \frac{3}{2} v_1^2 \right) \\ 
&- \frac{m_1}{2} \tilde \phi(\mathbf{y}_1)^2 + m_1 \tilde{\mathbf{A}}(\mathbf{y}_1) \cdot \mathbf{v}_1 + \frac{G_N m_1 m_2}{r} \tilde \phi(\mathbf{y}_1) \\
& + (1 \leftrightarrow 2) \; ,
\end{split}
\end{align}
where $ \mathcal{L}_0$ was introduced in Eq.~\eqref{eq:def_L0}, and the last term comes from the Feynman diagram with one external $\tilde \phi$ and one internal $\bar \phi$, represented in Figure~\ref{fig:feynDiagr_phitilde}.

%\paragraph{The Lagrangian up to 1PN order}

\subsection{Center-of-mass coordinates} \label{sec:CM}

Given the full 1PN two-body Lagrangian in Eq.~\eqref{eq:L1PN_unexpanded}, there remains to expand the two point-particle positions relatively to their common center-of-mass (CM). It is a well-known fact that there is no universal CM definition in General Relativity~\cite{1948RSPSA.195...62P}. For example, the ambiguities in the choice of the CM are related to the so-called "spin supplementary condition" for spinning point-particles, which is a gauge choice for the spin degree of freedom~\cite{HANSON1974498, Levi:2015msa}. We provide a discussion about the spin of our system and its relation to the center of mass in Appendix \ref{app:spin}. In our case, we will adopt the standard post-Newtonian definition of the CM, i.e at 1PN order:
\begin{align} \label{eq:CM_1PN}
\begin{split}
E \mathbf{Y}_\mathrm{CM} &= E_1 \mathbf{y}_1 + E_2 \mathbf{y}_2 \; , \\ \quad E_A &= m_A + \frac{1}{2} m_A v_A^2 - \frac{G_N m_1 m_2}{2 r} \; , \\ \quad E &= E_1 + E_2 \; .
\end{split}
\end{align}
Conversely, one can express the coordinates $\mathbf{y}_A$ using the relative separation $\mathbf{r}$ and the CM position $ \mathbf{X}_\mathrm{CM}$:
\begin{equation}\label{eq:x1x2_CM}
\mathbf{y}_1 =  \mathbf{Y}_\mathrm{CM} + (X_2 + \delta) \mathbf{r} \; , \quad \mathbf{y}_2 =  \mathbf{Y}_\mathrm{CM} + (-X_1 + \delta) \mathbf{r} \; ,
\end{equation}
where we have defined
\begin{align} \label{eq:def_CM_PN}
\begin{split}
X_A &= \frac{m_A}{m} \; , \quad m = m_1 + m_2 \; , \quad \mu =  \frac{m_1 m_2}{m} \; , \quad \nu = \frac{\mu}{m} \;
\end{split}
\end{align}
and to 1PN order we have:
\begin{equation}
	\quad \delta = - \nu \mathbf{V}_\mathrm{CM} \cdot \mathbf{v} + \nu (X_1 - X_2) \left( \frac{v^2}{2} - \frac{G_N m}{2r} \right).
\end{equation}

In the absence of any external field, the CM follows a straight line in the post-Newtonian coordinates. However, in the hierarchical three-body problem the binary CM will not follow such a trajectory even at the Newtonian level.

We now expand the Lagrangian~\eqref{eq:L1PN_unexpanded} in multipoles, e.g.
\begin{align}
\begin{split}
\tilde \phi( \mathbf{y}_1) &= \tilde \phi + (y_1 -  Y_\mathrm{CM})^i \partial_i \tilde \phi \\ &+ \frac{1}{2}  (y_1 -  Y_\mathrm{CM})^i (y_1 -  Y_\mathrm{CM})^j \partial_i \partial_j \tilde \phi + \dots \; ,
\end{split}
\end{align}
where the field is now evaluated at the CM position $ \mathbf{Y}_\mathrm{CM}$. The monopole corresponds to the term involving no derivatives of the fields, the dipole to the term involving first derivatives of the fields and so on.
% Note that contrary to the multipole expansion for the radiation field in the NRGR formalism, e.g CITE, the multipoles are not necessarily traceless since the fields do not need to be on-shell so that the trace part of the metric can indeed couple to the source.

\subsection{Osculating elements}

Before expanding the Lagrangian~\eqref{eq:L1PN_unexpanded} into multipoles and perform a matching with an effective point-particle action, we must eliminate an unwanted degree of freedom from the full theory. Indeed, we want to describe the evolution of the binary over a {\it secular} timescale, i.e. a time much longer than the period of the binary itself. In order to do so we average all quantities over the quick periodic motion of the binary, which can be approximated with an ellipse. Indeed, if the motion was purely Newtonian, the trajectory would be described by five constants of motion (six, if we count the initial time), which are nicely packaged in a set of geometrical elements. These are respectively the semimajor axis of the ellipse, the unit vector along the angular momentum and the Runge-Lenz vector:
\begin{align}
\begin{split}
a &= - \frac{G_Nm}{2} \left( \frac{v^2}{2} - \frac{G_N m}{r} \right)^{-1} \; , \\ \quad \hat{ \bm{\gamma}} &= \frac{\mathbf{r} \cross \mathbf{v}}{\sqrt{G_N ma(1-e^2)}} \; , \\ \quad \mathbf{e} &= \frac{1}{G_N m} \mathbf{v} \cross \left( \mathbf{r} \cross \mathbf{v} \right) - \frac{\mathbf{r}}{r} \; .
\end{split}
\end{align}
There are two angles in the unit vector $\hat{ \bm{\gamma}}$; furthermore $\mathbf{e}$ is orthogonal to $\hat{ \bm{\gamma}}$ (it points towards the perihelion) and its norm is equal to the eccentricity $e$. Conversely, the position and velocity vectors can be written as
\begin{align}
\begin{split} \label{eq:defOsculatingElements}
\mathbf{r} &= a \left( (\cos \eta - e) \; \hat{\bm{\alpha}} + \sqrt{1 - e^2} \sin \eta \; \hat{ \bm{\beta}}  \right) \; , \\
\mathbf{v} &= \sqrt{\frac{G_N m}{a}} \frac{1}{1 - e \cos \eta} \left( - \sin \eta \; \hat{\bm{\alpha}} + \sqrt{1 - e^2} \cos \eta \; \hat{ \bm{\beta}}  \right) \; ,
\end{split}
\end{align}
where $\hat{ \bm{\alpha}} = \mathbf{e} / e$, $\hat{ \bm{\beta}} = \hat{ \bm{\gamma}} \cross \hat{\bm{\alpha}} $ and $\eta$ is the eccentric anomaly, defined by
\begin{equation} \label{eq:def_eccentric_anomaly}
u = \sqrt{\frac{G_N m}{a^3}} t + \phi = \eta - e \sin \eta \; ,
\end{equation}
where $\phi$ is an arbitrary initial phase, and $u$ is called the mean anomaly.

Now, if the motion is slightly perturbed by post-Newtonian or quadrupolar corrections these constant elements will generically vary slowly with time (compared to the orbital frequency). Thus, in this generic case, we define the osculating elements as the (time-dependent) values of $a$, $e$, $\phi$, $\hat{ \bm{\alpha}}$ and $\hat{ \bm{\gamma}}$ such that the instantaneous position and velocity of the binary is given by the formulae~\eqref{eq:defOsculatingElements}. This physically corresponds to drawing at each point the ellipse defined by the instantaneous position and velocity of the binary. We have mapped the six components of $\mathbf{r}$, $\mathbf{v}$ into six elements $a$, $e$, $\phi$, $\hat{ \bm{\alpha}}$ and $\hat{ \bm{\gamma}}$. 

The equations of motion for the binary system can then be translated in a set of first-order equations on the osculating elements, called the Lagrange planetary equations (LPE). For completeness, we recall them in Appendix~\ref{app:LPE}. For our present purposes, though, it will be sufficient to state the result of Eq.\eqref{eq:kinetic_term_spin_app}, i.e. that the orbit-averaged LPE are completely equivalent to a spin kinetic term in the Lagrangian in flat spacetime:
\begin{equation} \label{eq:kinetic_term_spin}
 \frac{1}{2} \mu v^2 + \frac{G_N \mu m}{r} \rightarrow \bm{J} \cdot \bm{\Omega} \; ,
\end{equation}
where $\mathbf J$ is the total angular momentum of the binary and $\Omega$ is an angular velocity defined by
\begin{equation}\label{eq:def_J_Omega}
\bm{J} = \mu \sqrt{G_N ma(1-e^2)} \bm{\gamma} \; , \quad \bm{\Omega} =  \bm{\hat \alpha} \times \dot{\bm{\hat \alpha}} \; .
\end{equation}
Finally, we will average all quantities in the Lagrangian over one period of the binary, using the formula
\begin{equation}\label{eq:MeanValue_LowestOrder}
\left\langle A \right\rangle = \frac{1}{2 \pi} \int_0^{2 \pi} \mathrm{d} \eta (1 - e \cos \eta) A(\eta) \; .
\end{equation}
valid to lowest order for any quantity $A$ (we recall that $\eta$ is the eccentric anomaly defined in Eq.~\eqref{eq:def_eccentric_anomaly}). Thus, we will have removed from the Lagrangian the high-energy degree of freedom contained in the mean anomaly. As a consequence of the LPE~\eqref{eq:dot_a}, the semimajor axis $a$ will be conserved. This can be understood as deriving from the fact that the conserved conjugate momentum associated to the mean anomaly depends only on $ a $. As a side remark, notice that Eq.~\eqref{eq:MeanValue_LowestOrder} is valid only if we assume that the binary exactly follows an ellipse. As explained in App.~\ref{app:LPE}, there will be higher-order corrections to this formula, which however are not needed for our present purposes.

%To this aim, we introduce the osculating elements representing the ellipsis 

\section{Multipole expansion} \label{sec:multipolar}

\subsection{The internal Lagrangian} \label{sec:LInternal}

To begin with, let us deal with the very first term in the Lagrangian~\eqref{eq:L1PN_unexpanded}, namely the EIH Lagrangian. This term does not contain any coupling to the gravitational field. As explained before, it still contains  the short-distance degree of freedom from the Kepler trajectory of the binary system. In order to remove it and keep only the long-distance degrees of freedom which can be excited by the external field (in other words, the osculating elements), we should average the Lagrangian over the inner binary timescale, splitting the variables between the center-of-mass and the relative variables.

 \textit{A priori}, one should be careful about the fact that in the Newtonian kinetic energy one should use the relativistic center-of-mass definition in Eq.\eqref{eq:CM_1PN}. However, the meaning of the supplementary 1PN term will be better understood in terms of spin coupling, so we defer its calculation to a later Subsection. Thus, in this Subsection we stick to the Newtonian definition of the center-of-mass.  Carrying out the heavy but straightforward computations, we find by using Eq~\eqref{eq:MeanValue_LowestOrder}:
\begin{align}\label{eq:MeanValueLEIH}
\begin{split}
\left\langle \mathcal{L}_\mathrm{EIH}  - \mathcal{L}_0 \right\rangle &= \frac{1}{2} m V_\mathrm{CM}^2 +   \frac{1}{8} m V_\mathrm{CM}^4  \\ & + 3 \mu \frac{G_N^2 m^2}{a^2 \sqrt{1-e^2}} - \frac{G_N \mu m}{4 a} V_\mathrm{CM}^2 \; ,
\end{split}
\end{align} 
where we have dropped an unimportant constant term in the average (depending on the semimajor axis $a$ only, which is constant in the adiabatic approximation). Each term in Eq.~\eqref{eq:MeanValueLEIH} lends itself to a very simple interpretation. The two first terms are just the usual relativistic expansion of the center-of-mass velocity $-m \sqrt{1-V_\mathrm{CM}^2}$. The third term is the average of the EIH Lagrangian of a binary system in isolation: used in the LPE equation~\eqref{eq:dot_omega}, it gives rise to the celebrated perihelion precession formula. We call such a term the "internal" Lagrangian $\mathcal{L}_\mathrm{internal}$:
\begin{equation}
\mathcal{L}_\mathrm{internal}  = 3 \mu \frac{G_N^2 m^2}{a^2 \sqrt{1-e^2}} \; .
\end{equation}
Finally, the meaning of the last term in Eq.\eqref{eq:MeanValueLEIH} will become clearer in the next Subsection.

\subsection{Monopole} \label{sec:monopole}
Starting from Eq.\eqref{eq:L1PN_unexpanded} we can collect all the terms coupling the binary system to the monopole of the external gravitational field:
\begin{align}
\begin{split}
\mathcal{L}_{\mathrm{monopole}} &= -m \tilde \phi \left( 1 + \frac{3}{2} V_\mathrm{CM}^2 + \frac{3}{2} \nu v^2 - \frac{2 G_N \mu}{r} \right) \\ &- \frac{m}{2} \tilde \phi^2 + m \tilde{\mathbf A} \cdot \mathbf{V}_\mathrm{CM} \; ,
\end{split}
\end{align}
where $\tilde \phi$ and $\tilde{\mathbf A}$ are evaluated at the CM position $\mathbf{Y}_\mathrm{CM}$.
Note that in the term multiplying $\tilde \phi$ in the above equation, we have used the Newtonian version of the CM (i.e., we have set $\delta = 0$ in~\eqref{eq:x1x2_CM}) since the terms involving $\delta$ are of higher post-Newtonian order. Averaging over the binary orbital timescale, we find
\begin{align}\label{eq:RMonopole_LEIHunknown}
\begin{split}
\left\langle\mathcal{ L}_{\mathrm{monopole}} \right\rangle &= -m \tilde \phi \left( 1 + \frac{3}{2} V_\mathrm{CM}^2- \frac{ G_N \mu}{2a} \right) \\ &- \frac{m}{2} \tilde \phi^2 + m \tilde{\mathbf A} \cdot \mathbf{V}_\mathrm{CM} \; ,
\end{split}
\end{align}

Let us now gather this monopole coupling together with the average of the EIH Lagrangian~\eqref{eq:MeanValueLEIH} computed in the last Subsection. To 1PN order, we find
\begin{align}\label{eq:L3body_expanded}
\begin{split}
\left\langle\mathcal{ L}_{\mathrm{monopole + EIH}} \right\rangle  &= \mathcal{L}_\mathrm{internal} - m \sqrt{-\tilde g_{\mu \nu}V_\mathrm{CM}^\mu V_\mathrm{CM}^\nu } \\
&+ \frac{G_N \mu m}{2 a} \left( \tilde \phi - \frac{V_\mathrm{CM}^2}{2} \right) \; .
\end{split}
\end{align}
To 1PN order, the last term can be exactly accounted for by replacing the mass $m$ of the binary system (which is now treated as an effective point-particle) with its total energy in the worldline Lagrangian:
\begin{equation}\label{eq:Leff_with_E}
\left\langle\mathcal{ L}_{\mathrm{monopole+EIH}} \right\rangle  = \mathcal{L}_\mathrm{internal} - \mathcal{E} \sqrt{-\tilde g_{\mu \nu}V_\mathrm{CM}^\mu V_\mathrm{CM}^\nu } \; ,
\end{equation}
where $\mathcal E$ is defined as
\begin{equation}
 \mathcal{E} = m - \frac{G_N m \mu}{2a}  \; .
\end{equation}
Thus, the binary moves in the external field with a total mass equal to its binding energy, as could have been anticipated from an EFT perspective~\cite{Levi:2018nxp}. However, our computation highlights the fact that one should also include the internal Lagrangian in the effective action so that the binary PN precession effects are taken into account.

\subsection{Dipole} \label{sec:dipole}

Expanding the Lagrangian~\eqref{eq:L1PN_unexpanded} to dipole order (i.e, to first derivatives in the external fields) by taking into account the relativistic CM definition~\eqref{eq:CM_1PN}, we find at 1PN:
\begin{align}
\begin{split}
\mathcal{L}_\mathrm{dipole} &= \mu r^i \partial_i \tilde \phi \bigg[ - 2 \mathbf{V}_\mathrm{CM} \cdot \mathbf{v} \\
& + (X_1-X_2) \left(v^2 - \frac{G_N m}{2r} \right) \bigg] + \mu r^i v^j \partial_i \tilde A_j \; ,
\end{split}
\end{align}
where $X_A = m_A/m$. As before, one should average this Lagrangian over the inner binary timescale. We find that the term proportional to the difference of masses averages out, leaving us with an averaged Lagrangian
\begin{align} \label{eq:SpinCoupling}
\begin{split}
\left\langle\mathcal{ L}_{\mathrm{dipole}} \right\rangle &= \frac{\mu}{2} \sqrt{G_N m a (1-e^2)} \epsilon_{i j k} \hat \gamma^k \times \\
& \left(  2 V_\mathrm{CM}^i \partial^j \tilde \phi + \partial^i \tilde A^j  \right) \; .
\end{split}
\end{align}
From this expression one easily recognizes the coupling of a spinning point-particle to gravity given in e.g.~\cite{Levi:2018nxp,Levi:2015msa}. In our case, the spin tensor $J_{\mu \nu}$ depends on the total orbital angular momentum of the binary system $\bm{J} = \mu \sqrt{G_N m a(1-e^2)} \hat{\bm{\gamma}}$ through the relations
\begin{equation}\label{eq:Sij}
J_{ij} = \epsilon_{i j k} J^k \; , \quad J_{0i} = 0 \; .
\end{equation}
The second condition is called a spin supplementary condition, removing the unwanted degrees of freedom from the full spin tensor $J_{\mu \nu}$. As mentioned before, this gauge condition is related to the choice of a center-of-mass of the binary system; our particular CM choice in Eq.~\eqref{eq:CM_1PN} has selected the spin supplementary condition $J_{0i} = 0$, which has already been discussed e.g. in~\cite{10.2307/98894, Levi:2015msa}. We further elaborate on this in Appendix \ref{app:spin}. 
Furthermore, note that in Eq.~\eqref{eq:Sij} the spin tensor has been projected to a locally flat frame through $J^{ab} = e_\mu^a e_\nu^b J^{\mu \nu}$, where we have introduced the worldline tetrads defined over all spacetime by $\tilde g_{\mu \nu} e^\mu_a e^\nu_b = \eta_{ab}$. As a side remark, note that on top of the spin supplementary condition, the components of the spin vector are not all independent degrees of freedom following the remark below Eq.~\eqref{eq:def_J_Omega_app}. This reflects the fact that the spin of the inner binary contains two degrees of freedom once the orbital timescale has been integrated out, instead of the three degrees of freedom contained in the Euler angles of a generic spin.

To 1PN order, the spin coupling~\eqref{eq:SpinCoupling} can be written in a compact form using the Ricci rotation coefficients:
\begin{equation} \label{eq:SpinCoupling_compact}
\left\langle\mathcal{ L}_{\mathrm{dipole}} \right\rangle =   \frac{1}{2} J_{ab} \omega_{\mu}^{ab} V_\mathrm{CM}^\mu \; , \quad \omega_\mu^{ab} = e^{a\nu} D_\mu e_\nu^b \; .
\end{equation}
This formula gives back our previous equation~\eqref{eq:SpinCoupling} when expanded for weak-field values~\cite{Levi:2018nxp,Levi:2010zu}. We may be tempted to add to this spin coupling the kinetic term for the spin in Eq.~\eqref{eq:kinetic_term_spin} to obtain the minimal gravitational spin coupling which has been discussed at length in the NRGR formalism~\cite{Porto_2006, Levi:2015msa}:
\begin{equation}\label{eq:minimal_spin_coupling}
\mathcal{L}_\mathrm{spin} = \frac{1}{2} J_{\mu \nu} \Omega^{\mu \nu} \; .
\end{equation}
In this equation the total angular velocity $\Omega^{\mu \nu}$ includes both the Ricci rotation coefficients from Eq.~\eqref{eq:SpinCoupling_compact} and the locally flat angular velocity from Eq.~\eqref{eq:kinetic_term_spin}. It is defined through
\begin{equation}
\Omega^{\mu \nu} = e_a^\mu e_b^\nu \left( \Omega_\mathrm{flat}^{ab} + V_\mathrm{CM}^\alpha \omega_\alpha^{ab} \right) \; ,
\end{equation}
%\tilde e_A^\mu \frac{\mathrm{d} \tilde e^{A \nu}}{\mathrm{d}t} \; , \tilde{e}_A^\mu = \Lambda_A^a e_a^\mu
Here $\Omega_\mathrm{flat}^{ab}$ is related to the tensor $\Omega^{ij}=\epsilon_{ijk}\Omega^k$ by a relation that we discuss in Appendix~\ref{app:spin},
and the rotation vector $\bm{\Omega} = \hat{\bm{\alpha}} \times \dot{\hat{\bm{\alpha}}} $ has been defined in Eq~\eqref{eq:def_J_Omega}. However, there is a small piece that is still missing to obtain the full Eq.~\eqref{eq:minimal_spin_coupling}, related to the choice of the center-of-mass. As we show in Appendix \ref{app:spin}, in the spin gauge we are using ($J^{0i}=0$), there should be a supplementary spin kinetic term related to Thomas precession, which is 1PN order higher than the kinetic term~\eqref{eq:kinetic_term_spin}:
\begin{equation}\label{eq:spin_kinetic_term_1PN}
\frac{1}{2} J_{\mu \nu} \Omega^{\mu \nu} \supset \mathbf{J} \cdot \mathbf{\Omega} + \frac{1}{2} J_{ij} A_\mathrm{CM}^i V_\mathrm{CM}^j \; ,
\end{equation}
where $\mathbf{A}_\mathrm{CM}$ is the acceleration of the center-of-mass. Such a term is related to the PN corrections to the center-of-mass position (and speed) which we ignored in Section~\ref{sec:LInternal}. Indeed, using the full CM definition~\eqref{eq:CM_1PN} in the Newtonian part of the EIH Lagrangian~\eqref{eq:LEIH} gives a supplementary 1PN order term,
\begin{equation}
\mathcal{L}_\mathrm{Thomas} = m \mathbf{V}_\mathrm{CM} \cdot \frac{\mathrm{d}}{\mathrm{d}t} \left( \delta \mathbf{r} \right) \; , 
\end{equation}
where $\delta$ has been defined in Eq.~\eqref{eq:def_CM_PN}. At first sight, we may be tempted to discard such a term when averaging out the internal binary timescale. However, one should not forget to take also the time derivative acting on $\mathbf{V}_\mathrm{CM}$ in $\delta$, giving rise to
\begin{equation}\label{eq:Thomas}
\left\langle\mathcal{ L}_{\mathrm{Thomas}} \right\rangle = - \mu \left\langle r^i v^j \right\rangle V_\mathrm{CM}^i A_\mathrm{CM}^j = \frac{1}{2} J_{ij} A_\mathrm{CM}^i V_\mathrm{CM}^j \; ,
\end{equation}
which is exactly the additional spin kinetic term shown in Eq.~\eqref{eq:spin_kinetic_term_1PN}.

\subsection{Quadrupole}

From the EFT point of view, at 1PN quadrupolar order the couplings to gravity are contained in two non-minimal worldline operators~\cite{goldberger_gravitational_2010}:
\begin{align}
\begin{split} \label{eq:nonminimal_quadru}
\mathcal{O}_1 = \frac{1}{2} \int \mathrm{d} \tau E_{ij} I^{ij} \; , \quad
\mathcal{O}_2 = - \frac{4}{3} \int \mathrm{d} \tau B_{ij} J^{ij} \; ,
\end{split}
\end{align}
where $I^{ij}$ and $J^{ij}$ are the electric-type and magnetic-type quadrupole moments of the source, coupled to the corresponding parts of the Weyl tensor $C_{\mu \nu \alpha \beta}$:
\begin{align}
\begin{split}
E_{\mu \nu} &= C_{\mu \nu \alpha \beta} V_\mathrm{CM}^\alpha V_\mathrm{CM}^\beta \; , \\
B_{\mu \nu} &= \frac{1}{2} \epsilon_{\mu \alpha \beta \sigma} C^{\alpha \beta} {}_{\nu \rho} V_\mathrm{CM}^\sigma V_\mathrm{CM}^\rho \; .
\end{split}
\end{align}
Furthermore,in Eq.~\eqref{eq:nonminimal_quadru} the tensors have been projected to the locally flat frame defined below Eq.~\eqref{eq:Sij}.

%The electric-type quadrupole is often decomposed in an intrinsic part due to the finiteness of the extended body and a spin-induced moment, present only when the object is rotating~\cite{Levi:2018nxp,Levi:2015msa}. However, since our pointlike object is rotating by its very definition, such a separation is quite arbitrary in our case.

We could proceed as before and carry out the integration procedure to obtain the quadrupole moment of the effective point-particle. However, at this order the procedure is somewhat more involved than one could naively expect. The first complication comes from the corrections to the time averages introduced in Eq.~\eqref{eq:MeanValue_LowestOrder}. Indeed, post-Newtonian corrections to the period of the system will matter when taking the average of the Newtonian quadrupole moment, combining to produce a quadrupolar 1PN term. In the same way, the Newtonian quadrupolar corrections to the motion of the inner binary should be taken into account in the average of the EIH Lagrangian. 

The second complication comes from the corrections to the adiabatic approximation mentioned in the introduction. Indeed, in our analysis we are assuming that all the variables of the inner binary vary on long timescales (except of course the mean anomaly). This neglects short-timescale oscillations, which can ultimately have an effect on long-wavelength modes~\cite{Will:2020tri, Luo_2016}. It turns out that at lowest order this effect produces cross-terms of 1PN quadrupolar order~\cite{Lim:2020cvm, PhysRevD.89.044043} (no such corrections appear at lower multipole orders). While noting in passing that these kind of corrections have a very transparent meaning in the EFT language (they are high-energy corrections to an effective low-energy action), we will defer their complete calculation to further work.

\section{Integrating out the outer binary timescale} \label{sec:integrate_outer}

Now that we have replaced the binary system with an effective point-particle, we can integrate out the external fields $\tilde \phi$, $\tilde{ \mathbf{A}}$ in the presence of a third point-particle of mass $m_3$. For simplicity, in the following we will assume this mass to be of the same order of the mass of the inner binary: $ m_3\sim m $. 
%Thus, our results will be typically relevant for triple systems composed of $1-100 M_\odot$ BHs, which are quite common in dense star clusters~\cite{2020ApJ...903...67M} and whose mergers are observed by the LIGO/Virgo interferometers. 
 We will first derive the Feynman rules of the effective point-particle; then, in a second step, we will integrate out the outer binary timescale and comment on the different terms obtained in the expansion of the Lagrangian. For the 1PN precision we aim to, it will be sufficient to set the total Newtonian center-of-mass of the three-body system to the origin of coordinates (it will accelerate only at 2PN order~\cite{AIHPA_1985__43_1_107_0}). Thus, we will have the expressions
\begin{equation}
\mathbf{Y}_\mathrm{CM} =  X_3  \mathbf{R} \; , \quad  \mathbf{y}_3 = - X_\mathrm{CM} \mathbf{R} \; ,
\end{equation}
where we recall that $\mathbf{Y}_\mathrm{CM}$ is the position of the center-of-mass of the inner binary, and we have defined $ \mathbf{R} = \mathbf{Y}_\mathrm{CM} -  \mathbf{y}_3 $, $R = \vert \mathbf{R} \vert$, $\mathbf{N} = \mathbf{R}/R$, $M = m_1 + m_2 + m_3$, $X_3 = m_3 / M$ and $X_\mathrm{CM} = m/M$. The averages over the outer binary timescale are then taken in the same way than in the preceding Section.

\subsection{Power-counting rules}

Let us recap what we have learned so far and set up power-counting rules for the vertex coupling the binary system (now treated as an effective point-particle) to gravity. Up to dipole order, the Lagrangian of the binary system can be written as
\begin{equation} \label{eq:action_dipole_verySimple}
\mathcal{L} = \mathcal{L}_\mathrm{internal} - \mathcal{E} \sqrt{-\tilde g_{\mu \nu} V_\mathrm{CM}^\mu V_\mathrm{CM}^\nu} + \frac{1}{2} J_{\mu \nu} \Omega^{\mu \nu} \; .
\end{equation}
%where $\mathcal{L}_0$ is defined in Eq.~\eqref{eq:def_L0}, the second and third terms have been derived in Section~\ref{sec:monopole}, and the last term is the spin coupling from Section~\ref{sec:dipole}.
%In this equation we have introduced the "internal" Lagrangian, which is the average value of the Lagrangian (to 1PN order) in the absence of any external field:
%\begin{equation}
%\mathcal{L}_\mathrm{internal}  = \frac{G_N \mu m}{2a} + 3 \mu \frac{G_N^2 m^2}{a^2 \sqrt{1-e^2}} \; .
%\end{equation}
%where we have dropped the term $L \dot u$ as well as the Newtonian term $G \mu m / 2a$ from the LPE because we have averaged over the inner binary motion, so that $a$ and $L$ are constant and these terms do not contribute to the dynamics.% Also, note that from now on the averages are 
%we will drop the notation concerning the averages of the inner binary timescale, since we now have completely integrated it out.

Note that this Lagrangian has not yet been averaged over the period of the outer orbit $T_3$, and can therefore describe the secular dynamics on timescales shorter than $T_3$.
With such a simple Lagrangian, one can assign the standard power-counting rules of NRGR which have been described in e.g~\cite{goldberger_effective_2006, porto_effective_2016, Kuntz:2019zef}, considering the motion of the effective point-particle and the third mass (the outer orbit) for which one has $V_\mathrm{CM}^2 \sim v_3^2 \sim GM/a_3$ where   $a_3$ is the semimajor axis of the outer orbit. Thus, spatial derivatives are treated as $\partial_i \sim a_3^{-1}$. Time intervals scale as $t \sim a_3 / V_\mathrm{CM}$ and the metric perturbations scale as $\tilde \phi \sim \tilde A_i \sim V_\mathrm{CM}^{1/2} (\mpl a_3)^{-1}$. As usual in NRGR, the lowest-order Lagrangian scales as the orbital angular momentum of the outer orbit $J_3 \sim M V_\mathrm{CM} a_3$ which is treated non-perturbatively, higher-order corrections coming with higher powers of $V_\mathrm{CM}$.

However, one difference with respect to the standard NRGR power-counting rules is evidently the presence of two expansions, the first one in $v$ and the second one in $\varepsilon \equiv a/a_3$. \textit{A priori}, we could also have an expansion in the post-Newtonian parameter of the outer orbit $V_\mathrm{CM}$. However, not all these parameters are independent.
We choose to write all the post-Newtonian corrections as an expansion in the velocity of the inner binary $v$, converting the center-of-mass velocity by means of the relation $V_\mathrm{CM} \sim v \varepsilon^{1/2}$, which holds when $ m\sim m_3 $. In Table~\ref{table:power_counting} we give the power-counting rules of the monopole and dipole vertex which we computed in Sections~\ref{sec:monopole} and~\ref{sec:dipole}. The effect of post-Newtonian corrections on the dynamics of the system is highly non-trivial, as it can lead to suppression as well as enhancement of the Kozai-Lidov oscillations depending on the part of parameter space explored~\cite{Naoz_2013}; we expect that our power-counting scheme will help in discriminating between the different behaviours observed.

\begin{table}
\center
\begin{tabular}{|c|c|}
\hline
Operator & Rule  \\
\hline
$\displaystyle{\frac{1}{2} m V_\mathrm{CM}^2}$ & $J_3$ \vphantom{\bigg[} \\
\hline
$\displaystyle{- m \tilde\phi}$ & $J_3^{1/2}$  \vphantom{\bigg[} \\
\hline
$\displaystyle{m  \tilde{ \mathbf{A}} \cdot \mathbf{V}_\mathrm{CM}}$ & $J_3^{1/2} v \varepsilon^{1/2}$  \vphantom{\bigg[} \\
\hline
$\displaystyle{\frac{1}{8} m  V_\mathrm{CM}^4}$ & $J_3 v^2\varepsilon $  \vphantom{\bigg[} \\
\hline
$\displaystyle{- \frac{3}{2} m  \tilde \phi V_\mathrm{CM}^2}$ & $J_3^{1/2} v^2 \varepsilon$  \vphantom{\bigg[} \\
\hline
$\displaystyle{- \frac{1}{2} m  \tilde \phi^2}$ & $v^2 \varepsilon$  \vphantom{\bigg[} \\
\hline
$\displaystyle{ \frac{G_N \mu m}{2a}  \tilde \phi}$ & $J_3^{1/2} v^2$  \vphantom{\bigg[} \\
\hline
$\displaystyle{- \frac{G_N \mu m}{4a} V_\mathrm{CM}^2}$ & $J_3 v^2$  \vphantom{\bigg[} \\
\hline
$\displaystyle{ J_{ij} V_\mathrm{CM}^i \partial^j \tilde \phi}$ & $J_3^{1/2} v^2 \varepsilon^{3/2} $  \vphantom{\bigg[} \\
\hline
$\displaystyle{ \frac{1}{2}  J_{ij}  \partial^i \tilde A^j}$ & $J_3^{1/2} v \varepsilon  $  \vphantom{\bigg[} \\
\hline
$\displaystyle{\frac{1}{2}  J_{ij} A_\mathrm{CM}^i V_\mathrm{CM}^j}$ & $J_3 v^2 \varepsilon^{3/2}$  \vphantom{\bigg[} \\
\hline

\end{tabular}
\caption[Power-counting rules]{Power-counting rules for the vertices obtained by expanding the effective point-particle action~\eqref{eq:action_dipole_verySimple} up to 1PN order, with $J_3 = (G_N M^3 a_3)^{1/2}$,  $v^2 = Gm/a$ and $\varepsilon = a/a_3$. For convenience, the integral over time is not displayed, although it should be included to obtain a dimensionless rule.}
\label{table:power_counting}
\end{table}

Notice that the scaling of the spin is somewhat different than the one usually presented in NRGR~\cite{Levi:2010zu, Porto_2006}. Indeed, when taking compact objects as point-particles the spin is given as an order-of-magnitude by
\begin{equation}
J \sim m r_s v_\mathrm{rot} < m r_s \; ,
\end{equation}
where $v_\mathrm{rot}$ is the rotation velocity of the object and $r_s \sim G_N m$ its size. As a consequence, the ratio of the spin coupling presented in Eq.~\eqref{eq:SpinCoupling} to the Newtonian gravitational coupling is of $v^3$ (1.5PN) order. However, in our case the spin order-of-magnitude is given by $J \sim \mu \sqrt{G_N m a}$ so that the ratio of~\eqref{eq:SpinCoupling} to the Newtonian coupling is
\begin{equation}
\frac{J V_\mathrm{CM} \partial \phi}{m \phi} \sim v^2 \varepsilon^{3/2} \; .
\end{equation}
Thus, the inner binary angular momentum coupling is formally of 1PN order, although it is suppressed by the small ratio $\varepsilon^{3/2}$. This power-counting is different from the one of the Lense-Thirring precession caused by the intrinsic spin of the objects, which has been studied in~\cite{Liu_2020, Fang_2020} and enters at 1.5 PN.

%{\color{blue}This power-counting is somewhat different than the one presented in other studies of the hierarchical three-body problem~\cite{Liu_2020, Fang_2020}.}
%{(\color{blue} and not 1.5PN as claimed in~\cite{Liu_2020} ) }

\subsection{Monopole}

%Now that we have replaced the binary system with an effective point-particle, we can integrate out the external fields $\tilde \phi$, $\tilde{ \mathbf{A}}$ in the presence of a third point-particle of mass $m_3$. The resulting monopolar Lagrangian of the three-body system at 1PN order is remarkably simple: it is given by
Let us begin by integrating out the vertex contained in the monopole operators of the effective binary system, i.e in the square root appearing in Eq.~\eqref{eq:action_dipole_verySimple}. The final effective action, including orders of $J_3 v^2 \varepsilon$, is given by:
\begin{equation} \label{eq:Lv^2epsilon}
\mathcal{L}_{\leq v^2 \varepsilon}   = \mathcal{L}_\mathrm{internal} + \tilde{ \mathcal{L}}_\mathrm{EIH}^\mathrm{CM,3} \; ,
\end{equation}
where $  \mathcal{L}_\mathrm{EIH}^\mathrm{CM,3}$ is the EIH Lagrangian of the system composed by the CM (of mass $\mathcal{E}$, defined in Eq.~\eqref{eq:Leff_with_E}) and the third particle.

 % one replaces the mass of the binary by its binding energy $\mathcal{E}$.

%\begin{equation}
%\left\langle\mathcal{ R}_{\mathrm{monopole}} \right\rangle_I  = \mathcal{L}_\mathrm{internal} + \mathcal{L}_\mathrm{EIH}^\mathrm{CM,3}  - \frac{G_N \mu m}{2a} \left( \frac{V_\mathrm{CM}^2}{2} + \frac{G_N m_3}{R} \right) \; ,
%\end{equation}
%where $\mathcal{L}_\mathrm{EIH}^\mathrm{CM,3}$ is the EIH Lagrangian of the system composed by the CM (of mass $m$) and the third particle, with $\mathbf{R} = \mathbf{x}_3 - \mathbf{x}_\mathrm{CM}$, $R = \vert \mathbf{R} \vert$ and $\mathbf{N} = \mathbf{R}/R$. Or, using the nice understanding from Eq.~\eqref{eq:Leff_with_E}, we could also rewrite it as
%\begin{equation}
%\left\langle\mathcal{ R}_{\mathrm{monopole}} \right\rangle_I  = \mathcal{L}_\mathrm{internal} + \tilde{ \mathcal{L}}_\mathrm{EIH}^\mathrm{CM,3} \; ,
%\end{equation}
%where in $ \tilde{ \mathcal{L}}_\mathrm{EIH}^\mathrm{CM,3}$ one replaces the mass of the binary by its binding energy $\mathcal{E}$.

%Before moving on to the dipole terms, let us study the influence of the monopolar terms on the long-term evolution of the osculating elements of the binary system. Indeed, 
The Lagrangian in Eq.~\eqref{eq:Lv^2epsilon} involves a non-trivial coupling between the variables of the inner and outer binaries, given by
\begin{equation}
\mathcal{L}_{v^2} = - \frac{G_N \mu m}{2a} \left( \frac{V_\mathrm{CM}^2}{2} + \frac{G_N m_3}{R} \right) \; .
\end{equation}
This contribution is of order $ v^2 $ with respect to the standard Newtonian term $\mathcal{L}_0 \sim G_N M /a_3$. 
We average this term over one orbit of the outer binary, which gives
\begin{equation}
\mathcal{ L}_{v^2} = - \frac{G_N^2 M^2 \mu \nu_3}{2 a a_3} \left( 1 + \frac{X_3}{2} \right) \; ,
\end{equation}
where $M = m_3+m$, $X_3 = m_3/M$ and $\nu_3 = m m_3/M$.
This new monopole coupling has no effect on the dynamics. 
Indeed it depends only on the semimajor axes $a$ and $a_3$. Consequently, in the Lagrange planetary equations this term will only enter in the equation for the mean anomaly~\eqref{eq:dot_M}, which is irrelevant in the adiabatic approximation. Therefore, at the level of the monopole, the resulting motion is the one of two ellipses precessing because of standard two-body GR effects. 

In fact, one can be quite generic about the monopole terms. Indeed, the only planetary elements upon which the monopole terms could depend are the semimajor axes $a,a_3$ and the eccentricities $e,e_3$ (they do not involve angles). In the LPE the derivatives with respect to these elements enter only in the equations for the mean anomaly~\eqref{eq:dot_M} and the perihelion angle~\eqref{eq:dot_omega}. Thus, the only effect that monopole terms can have is to make the ellipses precess.

\subsection{Dipole}

\begin{figure}
	\centering
	\subfloat[]{
%		\begin{tikzpicture}
%			\begin{feynman}
%				\vertex (i1);
%				\vertex [right=of i1, dot] (a) {};
%				\vertex [right=of a] (f1);
%				\vertex [below=of i1] (i2);
%				\vertex [below=of a] (b);
%				\vertex [below=of f1] (f2);
%				
%				\diagram*{
%				(i1) -- (a) [dot] -- (f1),
%				(i2) -- (b) -- (f2),
%				(a) [dot] -- [scalar] (b),
%				};
%				
%				
%			\end{feynman}
%		
%		\end{tikzpicture}
\includegraphics[width=0.36\columnwidth]{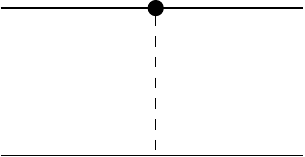}
	}\hspace{1em}
	\subfloat[]{
%		\begin{tikzpicture}
%			\begin{feynman}
%				\vertex (i1);
%				\vertex [right=of i1, dot] (a) {};
%				\vertex [right=of a] (f1);
%				\vertex [below=of i1] (i2);
%				\vertex [below=of a] (b);
%				\vertex [below=of f1] (f2);
%				
%				\diagram*{
%				(i1) -- (a) [dot] -- (f1),
%				(i2) -- (b) -- (f2),
%				(a) [dot] -- [ghost] (b),
%				};
%				
%				
%			\end{feynman}
%		
%		\end{tikzpicture}
\includegraphics[width=0.36\columnwidth]{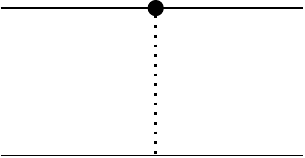}
	}\hspace{1em}

\caption{Feynman diagrams contributing to the lowest-order spin-orbit coupling, at  order $J_3 v^2 \varepsilon^{3/2}$. The dot represent the insertion of a spin coupling from Eq.~\eqref{eq:SpinCoupling}. The dotted line represents propagation of a scalar $\phi$, while the dashed line stands for the propagation of a vector $\mathbf A$.}
\label{fig:feynDiagr_spinOrbit}
\end{figure}

In order to integrate out modes contributing to the potential at dipole order, we have to compute the diagrams related to spin-orbit coupling. These are shown in Figure~\ref{fig:feynDiagr_spinOrbit}. Using the Lagrangian averaged over the inner orbit Eq.~\eqref{eq:SpinCoupling}, we find:
\begin{equation}\label{eq:spinorb}
\mathcal{L}_\mathrm{spin-orbit} = \frac{1}{2} J_{ij} \frac{G_N m_3}{R^3} R^i \left(4 v_3^j - 2 V_\mathrm{CM}^j \right) \; .
\end{equation} 
At this order of approximation however, we should also take into account the Thomas precession term of Eq.~\eqref{eq:Thomas}. This gives a contribution of the same size of the terms in Eq.\eqref{eq:spinorb}. We can replace the center-of-mass acceleration in Eq.~\eqref{eq:Thomas} using the equation of motion, since the difference between the two terms would contribute at a higher PN order (this is usually called the "double zero trick"~\cite{SCHAFER1984128, doi:10.1063/1.529135, BARKER1980231}). Thus, at order $J_3 v^2 \varepsilon^{3/2}$ the full Lagrangian is given by
\begin{align}
\begin{split}
\mathcal{L}_{ v^2 \varepsilon^{3/2}} &= \frac{1}{2} J_{ij} \frac{G_N m_3}{R^3} R^i \left(4 v_3^j - 3 V_\mathrm{CM}^j \right)\\
& = - \frac{1}{2} J_{ij} \frac{G_N m_3 ( 4m + 3 m_3)}{M R^3} R^iV^j \; ,
\end{split}
\end{align}
which recovers the result already known in the NRGR approach~\cite{Levi:2010zu}.
Carrying out the average over the outer binary timescale in a way very similar to the previous Section, we find
\begin{equation}
\left\langle \mathcal{L}_{ v^2 \varepsilon^{3/2}} \right\rangle = - \frac{4m + 3 m_3}{2 m} \frac{G_N }{a_3^3 (1-e_3^2)^{3/2}} \; \mathbf{J} \cdot \mathbf{J}_3 \; ,
\end{equation}
where $ \mathbf{J}_3$ is the angular momentum vector of the outer orbit, $ \mathbf{J}_3 = \mu_3 (G_N M a_3 (1-e_3^2))^{1/2}  \hat{\bm{\gamma}}_3$ (here $\hat{\bm{\gamma}}_3$ is the unit vector along the outer orbit angular momentum, and $\mu_3 = m m_3 / M$). Thus, this term is indeed a coupling between the angular momentum vectors of the two orbits.

From this expression one can obtain a precession equation for the inner orbit angular momentum. Indeed, varying the kinetic term for the spin with respect to the canonical variables $\hat{\bm{\alpha}}$ and $\bm{W} = \bm{J} \times \hat{\bm{\alpha}}$, one obtains the equations of motion
\begin{align}
\frac{\mathrm{d}\bm{W}}{\mathrm{d}t} &= - \Omega_\mathrm{prec} \bm{W} \times \bm{J}_3 \; , \\ \frac{\mathrm{d}\hat{\bm{\alpha}}}{\mathrm{d}t} &= \Omega_\mathrm{prec}  \bm{J}_3 \times \hat{\bm{\alpha}} \; ,
\end{align}
where the precession frequency is equal to
\begin{equation}
 \Omega_\mathrm{prec} = \frac{4m + 3 m_3}{2 m} \frac{G_N }{a_3^3 (1-e_3^2)^{3/2}} \; .
\end{equation}
From these two equations, and using the Jacobi identity for the cross product, one obtains the precession equation
\begin{equation}
\frac{\mathrm{d}\bm{J}}{\mathrm{d}t} =  \Omega_\mathrm{prec} \bm{J}_3 \times \bm{J} \; ,
\end{equation}
which is in complete accordance with earlier results on the hierarchical three-body problem~\cite{2019ApJ...883L...7L, Lim:2020cvm}. Notice that conservation of the total angular momentum requires that $\bm{J}_3$ satisfies an analogous equation,
\begin{equation}
\frac{\mathrm{d}\bm{J}_3}{\mathrm{d}t} =  \Omega_\mathrm{prec} \bm{J} \times \bm{J}_3 \; .
\end{equation}

 In particular, it was shown that this angular momentum precession may play an important role for stellar-mass binary mergers near a supermassive BH~\cite{2019ApJ...883L...7L}. Quadrupolar terms would lead to further precession effects, of order $J_3 v^2 \varepsilon^2$ in the Lagrangian. We leave the computation and the astrophysical implications of such terms to further work.

\section{Conclusions}

The NRGR approach to the two-body problem was designed to deal with extended compact objects. In this article, we have extended NRGR to the setting of a hierarchical three-body problem. In the approximation that the inner orbit is much smaller in amplitude than the outer one, the inner binary system can be replaced by an effective point-particle endowed with multipole moments, which we explicitly computed up to dipole order. This is very natural from the EFT perspective and provides a new specific example of how an extended (and not so compact) system can be accounted for by means of a point-particle operator.

Our procedure consists in integrating out the short timescales associated with the period of the two hierarchical orbits. One notable result of our study is to make explicit the link between the Lagrange planetary equations, describing the long-time evolution of the inner binary Keplerian parameters, and the kinetic term for a spin in the EFT language. We have also clarified the relation between the post-Newtonian definition of the center-of-mass and the spin supplementary condition for the angular momentum of the inner binary. The computation of quadrupolar post-Newtonian terms including the corrections to the adiabatic approximation will be the subject of a future publication.

Our study moves towards a more systematic characterization of the relativistic hierarchical three-body problem. Indeed, the EFT techniques that we employed can be applied to study efficiently three-body trajectories to higher orders in both PN and multipole expansions. Two immediate fields of application will be the study of the influence of relativistic three-body interactions on the Kozai-Lidov mechanism, and the production of three-body analytic waveforms in the PN regime using the effective two-body description. Another interesting follow-up would be to obtain (in a matching procedure) the multipole structure of the inner binary system to higher orders from a numerical relativistic three-body solver. Finally, while we have restricted here the discussion to objects of similar mass, it could also be interesting to generalize our work to the case where $m_3 \gg m_1,m_2$, which is particularly relevant for binary BHs orbiting a supermassive BH at the core of a galaxy. While it would be very easy to include this new large parameter in the power-counting rules in Table~\ref{table:power_counting}, it would probably be more efficient to take advantage of the large ratio of masses in order to explore the interplay between BH perturbation theory (for the outer orbit) and post-Newtonian EFT treatment (for the inner orbit).
 We plan to explore these avenues in a near future.

\section*{Acknowledgements}

We thank Vitor Cardoso, Michèle Levi and Leong Khim Wong for useful comments on the draft. ET thanks the participants of the KITP program “Probing Effective Theories of Gravity in Strong Fields and Cosmology” for stimulating discussions. This research was partly supported by the Italian MIUR under contract 2017FMJFMW (PRIN2017) and by the National Science Foundation under Grant No. NSF PHY-1748958.

\appendix

\section{Lagrange planetary equations and secular approximation} \label{app:LPE}

This Appendix introduces a set of equations initially introduced by Lagrange. 
To begin with, note that the time-dependence of the osculating elements defined by Eq.~\eqref{eq:defOsculatingElements} cannot be arbitrary. We must impose a gauge-fixing condition such that the velocity is indeed given by Eq.~\eqref{eq:defOsculatingElements}. We denote such a condition by
\begin{equation}
\mathbf{C} =  \frac{\mathrm{d} \mathbf{r}}{\mathrm{d}t}  = \mathbf{v} \; ,
\end{equation}
where the expression for the vector $v$ was given in Eq.~\eqref{eq:defOsculatingElements}. Thus, there is a relation between time derivatives of the osculating elements. This gauge-fixing condition removes three degrees of freedom (equivalently, six variables in phase space) from the six degrees of freedom contained in the six osculating elements (equivalently, twelve variables in phase space).

Now, we could write a Lagrangian for the osculating elements by implementing this constraint with a Lagrange multiplier $\bm{\lambda}$, so that
\begin{equation}
\mathcal{L} =  \frac{1}{2} \mu v^2 + \frac{G \mu m}{r} + \bm{\lambda} \cdot \big ( \mathbf{C} - \mathbf v \big) + \mathcal{L}_1 \; ,
\end{equation}
where $\mathcal{L}_1$ has been defined in Eq~\eqref{eq:def_L0}.
From there one could deduce the Lagrange planetary equations (LPE) which relate time derivatives of the osculating elements to the perturbing function $\mathcal{L}_1$. However, it is much easier to derive them in a Hamiltonian formalism, see e.g.~\cite{valtonen_karttunen_2006} to which we refer the reader interested in the details of the derivation.
% we find it much easier to assume that the LPE are known (they are simpler to derive in an Hamiltonian formalism), and write a Lagrangian which reproduces them.

The LPE are traditionally expressed using the following angles: $\iota$ is the inclination, $\omega$ the argument of periapsis, and $\Omega$ the longitude of the ascending node. In term of these, the unit vectors  $\hat{ \bm{\alpha}}$ and $\hat{ \bm{\gamma}}$ are expressed as
\begin{align} \label{eq:param_e_l}
\begin{split}
\hat{ \bm{\alpha}} &= R_z(\Omega) R_x(\iota) R_z(\omega) \hat{ \mathbf{u}}_x \; , \\ \quad\hat{ \bm{\gamma}} &= R_z(\Omega) R_x(\iota) R_z(\omega) \hat{ \mathbf{u}}_z \; ,
\end{split}
\end{align}
where $\hat{ \mathbf{u}}_x$, $\hat{ \mathbf{u}}_y$, $\hat{ \mathbf{u}}_z$ are the Cartesian basis vectors. Using these angles, the LPE are given by~\cite{valtonen_karttunen_2006}
\begin{align}
\dot a &= \sqrt{\frac{4a}{G_N m}} \frac{\partial  \tilde{\mathcal{L}_1}}{\partial u} \; , \label{eq:dot_a} \\
\dot e &= - \sqrt{\frac{1-e^2}{G_N m a e^2}} \frac{\partial  \tilde{\mathcal{L}_1}}{\partial \omega} + \frac{1-e^2}{\sqrt{G_N m a}e} \frac{\partial  \tilde{\mathcal{L}_1}}{\partial u} \; , \label{eq:dot_e} \\
\dot \iota& = - \frac{1}{\sqrt{G_N m a(1-e^2)} \sin \iota} \frac{\partial  \tilde{\mathcal{L}_1}}{\partial \Omega} \nonumber  \\
&+ \frac{\cos \iota}{\sqrt{G_N m a(1-e^2)} \sin \iota} \frac{\partial  \tilde{\mathcal{L}_1}}{\partial \omega} \; , \\
\dot u &= \sqrt{\frac{G_N m}{a^3}} - \sqrt{\frac{4a}{G_N m}} \frac{\partial  \tilde{\mathcal{L}_1}}{\partial a} - \frac{1-e^2}{\sqrt{G_N m a}e} \frac{\partial  \tilde{\mathcal{L}_1}}{\partial e} \; , \label{eq:dot_M} \\
\dot \omega &= \sqrt{\frac{1-e^2}{G_N m a e^2}} \frac{\partial \tilde{ \mathcal{L}_1}}{\partial e} - \frac{\cos \iota}{ \sqrt{G_N m a(1-e^2)} \sin \iota} \frac{\partial \tilde{ \mathcal{L}_1}}{\partial \iota} \; , \label{eq:dot_omega} \\
\dot \Omega &=  \frac{1}{\sqrt{G_N m a(1-e^2)} \sin \iota} \frac{\partial \tilde{\mathcal{L}_1}}{\partial \iota} \; ,
\end{align}
where $ \tilde{\mathcal{L}_1} = \mathcal{L}_1 / \mu$.
 It can be easily checked that the LPE can be derived from the following fist-order Lagrangian~\footnote{A subtlety can arise in the fact that the osculating elements are not appropriate if the perturbing function depends on $\mathbf{v}$ and should be replaced by the so-called \textit{contact elements}, see e.g~\cite{relativistic_celestial_mechanics}. However, in our case the difference in the resulting equations will be of 2PN order so that we do not have to worry about this. } 
\begin{equation}\label{eq:Lagrangian_planetary_angular}
\mathcal{L} = \mu \left[ \frac{G_N m}{2 a} + L \dot u + G \dot \omega + H \dot \Omega \right] + \mathcal{L}_1 \; ,
\end{equation}
The conjugate momenta are given by
\begin{equation}
L = \sqrt{G_N ma}  \; , \; G = L \sqrt{1-e^2} \; , \; H = G \cos \iota \; .
\end{equation}

This Lagrangian has the nice property to be exact (no secular approximation has been done, it is completely equivalent to the original Lagrangian~\eqref{eq:def_L0}). However, it is not manifestly invariant under a rotation of the basis vectors; such a manifest invariance can be recovered by noticing that the angular part can be rewritten as  
\begin{equation} \label{eq:kinetic_term_spin_app}
 \mu \left[ G \dot \omega + H \dot \Omega \right] = \mu G \; \hat{\bm{\beta}} \cdot \dot{\hat{\bm{\alpha}}} = \bm{J} \cdot \bm{\Omega} \; ,
\end{equation}
where $\mathbf J$ is the total angular momentum of the binary and $\Omega$ is an angular velocity defined by
\begin{equation} \label{eq:def_J_Omega_app}
\bm{J} = \mu \sqrt{G_N ma(1-e^2)} \hat{\bm{\gamma}} \; , \quad \bm{\Omega} =  \bm{\hat \alpha} \times \dot{\bm{\hat \alpha}} \; .
\end{equation}
Thus, the angular kinetic term can be identified with a spin coupling in flat space (note that our sign convention for the metric is different from the one used in e.g Refs~\cite{Levi:2018nxp,Levi:2015msa}, which explains the sign difference of the kinetic term). 
However, note that not all the components of the spin vector are independent, since the Lagrangian shown in~\eqref{eq:kinetic_term_spin_app} displays only two degrees of freedom (corresponding to four equations in phase space once a variational principle is applied).
Indeed, notice that if one wants to vary the Lagrangian with respect to $\hat{\bm{\alpha}} $ and $\hat{\bm{\beta}} $ in order to keep a manifest rotational invariance, one should also impose that these vectors should be unitary and orthogonal in order to preserve the right number of degrees of freedom.
% In other words, the six equations describing the evolution of a spin in phase space contain one unwanted degree of freedom that one can remove using the two supplementary conditions $\bm{\alpha} \cdot \bm{\gamma} = 0$ and $\bm{\alpha}^2 = 1$ (the norm of $\bm{J}$ being related to the eccentricity $e$). 

Finally, the LPE are often averaged over the periodic motion of the binary system: this is called the adiabatic or secular approximation. This corresponds to eliminating the short-distance degree of freedom contained in the mean anomaly $u$; as a consequence, since the perturbing function does not depend on $u$ any more, the semimajor axis $a$ is constant through time from Eq.~\eqref{eq:dot_a}. Thus, the two-body Lagrangian shown in Eq.\eqref{eq:Lagrangian_planetary_angular} is indeed equivalent to a spin kinetic term, since the term $G_N m/2a + L \dot u$ becomes an irrelevant constant in the adiabatic approximation.

 After this elimination, the binary system is described by four dynamical quantities (the eccentricity $e$ and the three Euler angles defined above) which vary over a timescale much greater than the period of the binary.
Technically, we use the formula valid for any quantity of interest $A$:
\begin{align} \label{eq:MeanValue}
\begin{split}
\left\langle A \right\rangle &= \frac{1}{T}\int_0^{T} \mathrm{d} t  \; A(t) = \frac{1}{T} \int_0^{2 \pi}  \frac{\mathrm{d}t}{\mathrm{d}\eta} \mathrm{d} \eta \; A(\eta) \; , \\ \quad T &=  \int_0^{2 \pi} \; \frac{\mathrm{d}t}{\mathrm{d}\eta} \mathrm{d} \eta \; .
\end{split}
\end{align} 
%where the $I$ subscript indicates that we average over the inner binary timescale (the same procedure will be used for the outer binary).
 Using Eq.~\eqref{eq:def_eccentric_anomaly}, one has
\begin{equation}
\frac{\mathrm{d}t}{\mathrm{d}\eta} = \frac{1-e\cos \eta}{\dot u + \dot e \sin \eta} \; .
\end{equation}
At lowest order in the perturbing function $\mathcal{L}_1$, one has $\dot u + \dot e \sin \eta = \sqrt{G_N m/a^3}$ and $T = 2 \pi \sqrt{a^3/(G_N m)}$, so that the mean value becomes
\begin{equation}\label{eq:MeanValue_LowestOrder_app}
\left\langle A \right\rangle = \frac{1}{2 \pi} \int_0^{2 \pi} \mathrm{d} \eta (1 - e \cos \eta) A(\eta) \; .
\end{equation}
 However there will be higher-order corrections to these quantities as is implied by Eqs.~\eqref{eq:dot_e}-\eqref{eq:dot_M}. These corrections contribute at the quadrupolar 1PN level, which is beyond the scope of this paper. They will be investigated in more details in a forthcoming publication, along with the corrections to the adiabatic approximation.
 % These corrections will be mentioned in Section~\ref{sec:beyond_secular}, where we will also compute the corrections to the adiabatic approximation coming from short-timescale fluctuations in the parameters of the binary.

\section{Spin kinetic term and gauge fixing of rotational variables} \label{app:spin}
In this appendix we provide some details of the computation of the spin kinetic term \eqref{eq:minimal_spin_coupling} as a function of the intrinsic angular momentum of the inner binary. The computations are analogous to those carried out in~\cite{Levi:2015msa}, with the difference that we specialize to the \textit{no mass dipole} gauge in which the time components of the spin tensor are set to zero. This choice will make simple to connect the spin tensor to the orbital angular momentum.

First of all, it is useful to introduce a worldline tetrad $ e^{\mu}_A(\sigma) $ defined only on the worldline $ y^{\mu}(\sigma) $ ($ \sigma $ being the affine parameter of the curve) which represents a choice of axes in the rest-frame of the body and satisfies: $  g_{\mu\nu}(y(\sigma)) e^{\mu}_A(\sigma)e^{\nu}_B(\sigma)=\eta_{AB} $. This tetrad can be used to define the angular velocity of the body: $ \Omega^{\mu\nu}=e^{\nu}_A (D e^{\mu A}/D\sigma) $, whose conjugate is the spin tensor $J_{\mu\nu}=2\partial\mathcal{L}/\partial\Omega^{\mu\nu}$. Both these tensors contain gauge degrees of freedom, since only the spatial orientation of the worldline tetrad has a physical meaning. In fact we can choose arbitrarily its time-like direction, encoded in $ e^{\mu}_{[0]} $. This gauge choice corresponds to a redundant boost transformation of the worldline tetrad 
(in order to avoid ambiguities between the different set of indices, we are using square brackets to distinguish the flat indices of the worldline tetrad from the others).

The gauge fixing of $ e^{\mu}_{[0]} $ must be supplemented with a gauge fixing of the conjugate variables in $ J_{\mu\nu} $, the so-called Spin Supplementary Condition (SSC). Starting from a covariant gauge choice in which $ e^{\mu}_{[0]}=p^{\mu}/\sqrt{-p^2} $ and the spin tensor satisfies the covariant SSC $ J_{\mu\nu}p^{\nu}=0 \,$, the action of a boost will change the worldline tetrad and the angular velocity tensor. The changes produced by this transformation in the Lagrangian can be interpreted by means of a redefinition of the spin tensor and a consequent change of the SSC. We will therefore use this boost degree of freedom to first pick the \textit{no mass dipole} SSC for the spin tensor and then to fix the canonical gauge for the angular velocity vector. In this way we will get an expression dependent only on the intrinsic angular momentum of the binary.

As computed in~\cite{Levi:2015msa}, the transformation of the spin kinetic term of the Lagrangian under a boost of the worldline tetrad (starting from the covariant gauge and SSC) is the following:
\begin{align}\label{eq:boostedspinkin}
	\dfrac{1}{2}J_{\mu\nu}\Omega^{\mu\nu}=\dfrac{1}{2}\hat{J}_{\mu\nu}\hat{\Omega}^{\mu\nu}+\dfrac{p^{\lambda}}{-p^2}\hat{J}_{\mu\lambda}\dfrac{Dp^{\mu}}{D\sigma}\,,
\end{align}
where we have used hatted symbols to label boosted variables and in particular we have defined the boosted spin tensor to be $ {J}_{\mu\nu}=\hat{J}_{\mu\nu}- \delta z_{\mu}p_{\nu}+\delta z_{\nu}p_{\mu}\,$, with $ \delta z_{\mu}=\hat{J}_{\mu\rho}p^{\rho}/(-p^2) $.
%$ {J}_{\mu\nu}=\hat{J}_{\mu\nu}- \dfrac{\hat{J}_{\mu\rho}p^{\rho}p_{\nu}}{p^2}+ \dfrac{\hat{J}_{\nu\rho}p^{\rho}p_{\mu}}{p^2}$.
We can interpret this change of the spin tensor as due to a shift of the center of the body rotation, that is the point where the worldline intersects the body. In the case of the \textit{no mass dipole} gauge, in which the spin tensor is purely spatial, this shift corresponds to setting the center of the worldline on the relativistic center of mass, as shown in the main text.
The second term in Eq.\eqref{eq:boostedspinkin} will instead contribute to the Thomas precession, which we can understand as due to a gravitational torque associated to the finite size of rotating objects in GR.\\

Before specifying the boost needed to get to the desired SSC, it is useful to disentangle the gravitational field from the spinning degrees of freedom. We can do so by introducing the gravitational tetrad field $ g_{\mu\nu}(x)\tilde{e}^{\mu}_a(x)\tilde{e}^{\nu}_b(x)=\eta^{ab} $, which is defined on the whole space-time. This tetrad can be related to the worldline tetrad by means of a Lorentz transformation: $ \tilde{e}^{\mu}_a(y(\sigma))=\Lambda_a^A(\sigma)e^{\mu}_A(\sigma) $, being  $  \Lambda_a^A(\sigma) $ a Lorentz matrix dependent on the affine parameter of the worldline. As for the worldline tetrad flat indices, when needed we will use round brackets to distinguish the flat indices of the tetrad field from the others. 

In this notation, once the gauge of the tetrad field is fixed, we can fix the time-like vector of the worldline tetrad by choosing the boosted zero components of the Lorentz matrices: $ \hat{\Lambda}_a^{[0]} $. Moreover, introducing the tetrad field will make possible to write all the objects in the right hand side of Eq.\eqref{eq:boostedspinkin} in terms of their counterparts with flat indices. Such quantities correspond to those computed in terms of the intrinsic angular momentum of the binary, as they are independent on the external gravitational field. In particular we have:
\begin{align}\label{eq:flatboosted}
	\dfrac{1}{2}\hat{J}_{\mu\nu}\hat{\Omega}^{\mu\nu}=\dfrac{1}{2}\hat{J}_{ab}\hat{\Omega}_{flat}^{ab}+\dfrac{1}{2}\hat{J}_{ab}\omega_{\mu}^{ab}u^{\mu}\,,
\end{align}
where $ \omega_{\mu}^{ab}=\tilde{e}^{a}_{\nu}\nabla_{\mu}\tilde{e}^{b\nu} $ is the spin connection of the tetrad field, $ u^{\mu}=d y^{\mu}/d\sigma $ is the worldline speed and we have defined  $ \hat{\Omega}_{flat}^{ab} = \hat{\Lambda}^b_A d\hat{\Lambda}^{aA}/d\sigma$.\\

At this point we can fix the gauge boost of the worldline tetrad. In order to set the time components of the spin tensor to zero, $ \hat{J}_{a(0)}=0 \,$, we need to choose a boost such that $ \sqrt{p^2}\hat{\Lambda}_{[0]a}= 2p_0 \delta_{0a}-p_a$ (this can be understood by inspecting the generic expression for $ \hat{J}_{\mu\nu} \,$, as discussed in~\cite{Levi:2015msa}). Doing so, we obtain the following:
\begin{align}\label{eq:nomassdip}
\dfrac{1}{2}J_{\mu\nu}\Omega^{\mu\nu}=&\dfrac{1}{2}\hat{J}_{(i)(j)}\hat{\Omega}_{flat}^{(i)(j)}+\dfrac{1}{2}\hat{J}_{(i)(j)}\omega_{\mu}^{(i)(j)}u^{\mu}\\\nonumber
&+\dfrac{p^{(j)}}{-p^2}\hat{J}_{(i)(j)}\tilde{e}^{(i)}_{\mu}\dfrac{Dp^{\mu}}{D\sigma},	
\end{align}
This gauge choice makes possible to unpack $ \hat{\Omega}_{flat}^{(i)(j)} $ and express $ \hat{\Lambda}^a_{[0]}d\hat{\Lambda}^{b[0]} / d\sigma $ in terms of the momentum of the worldline, leaving to compute only on the spatial part of the Lorentz matrices. However, these spatial leftovers won't be $ SO(3) $ matrices, since they need to satisfy the condition $ \hat{\Lambda}^a_A\eta^{AB}\hat{\Lambda}^b_B=\eta^{ab} $ and will carry a dependence on the worldline momentum, due to the gauge condition on $ \hat{\Lambda}^a_{[0]} $ .

In order to obtain an angular velocity tensor defined in terms of rotation matrices and to remove its dependence on the worldline momentum, we can take a further boost of the worldline tetrad. This time however, we will not use a redefinition of the spin tensor to absorb the new terms appearing in the Lagrangian after the transformation. Rather, we will retain the spin tensor satisfying the \textit{no mass dipole} SSC and we will keep track of the new terms explicitly.

In order to make $ \hat{\Lambda}^a_A $ an $ SO(3) $ matrix, we need to choose a gauge in which $ \hat{\Lambda}^a_{[0]}=\delta^a_0 $. Therefore we implement a boost of the worldline tetrad that sends the time-like unit vector $ (2p_0 \delta_{0}^a-p^a)/\sqrt{-p^2} $ to $ \delta^a_0 $ . This transformation will change only the first term in Eq.\eqref{eq:nomassdip} as follows:
\begin{align}
\dfrac{1}{2}\hat{J}_{(i)(j)}\hat{\Omega}_{flat}^{(i)(j)}=\dfrac{1}{2}\hat{J}_{(i)(j)}{\Omega}_{SO(3)}^{(i)(j)}+\dfrac{1}{2}\hat{J}_{(i)(j)}u^{(i)}\dfrac{du^{(j)}}{d\sigma}\,,
\end{align}
where now $ {\Omega}_{SO(3)}^{(i)(j)} $ is build out of rotation matrices and we have used $ p^a=m u^a/\sqrt{-u^2} $, with $ -u^2= 1 $ at leading order in the PN expansion.

Having fixed the gauge for both angular velocity and spin tensor, we can carry out the explicit computation of the last two terms in Eq.\eqref{eq:nomassdip}. In order to do so, we pick the tetrad field in such a way to have $ \tilde{e}^{(i)}_0=0 $. Then, at 1PN order we obtain:
\begin{align}
	\dfrac{1}{2}\hat{J}_{ab}\omega_{\mu}^{ab}u^{\mu} &=\dfrac{1}{2}\hat{J}_{(i)(j)}(4u^{(i)}\partial^{(j)}\tilde{\phi}+\partial^{(i)}\tilde{A}^{(j)})\; , \\\nonumber
	\dfrac{p^{(j)}}{-p^2}\hat{J}_{(i)(j)}\tilde{e}^{(i)}_{\mu}\dfrac{Dp^{\mu}}{D\sigma}&=\hat{J}_{(i)(j)}u^{(j)}\left(\dfrac{du^{(i)}}{d\sigma}+\partial^{(i)}\tilde{\phi}\right)\, .
\end{align}
Plugging these results into Eq.\eqref{eq:boostedspinkin}, and identifying the worldline with the trajectory of the center of mass, $ u^{(i)}=V_{CM}^{(i)} $ we finally get:
\begin{align}\label{eq:finalspin}
	\dfrac{1}{2}J_{\mu\nu}\Omega^{\mu\nu} = &\dfrac{1}{2}\hat{J}_{(i)(j)}{\Omega}_{SO(3)}^{(i)(j)}+\dfrac{1}{2}\hat{J}_{(i)(j)}A_{CM}^{(i)}V_{CM}^{(j)}\\\nonumber&+\dfrac{1}{2}\hat{J}_{(i)(j)}(2V_{CM}^{(i)}\partial^{(j)}\tilde{\phi}+\partial^{(i)}\tilde{A}^{(j)}) \; .
\end{align}
Then, with a mild abuse of notation, we can drop the index brackets and the hats so as to match the expressions used (for simplicity) in the main text: $ \hat{J}_{(i)(j)}\mapsto J_{ij}\;,\;{\Omega}_{SO(3)}^{(i)(j)}\mapsto \Omega^{ij} $. We stress however that these are different from the $ (\mu ,\nu)=(i,j) $ components of $ J_{\mu\nu} $ and $ \Omega_{\mu\nu} $, which depend on the external gravitational field.

Using this notation and the definitions $ J_{ij}=\epsilon_{ijk} J^k\;,\;\Omega_{ij}=\epsilon_{ijk}\Omega^k$, we can rewrite Eq.\eqref{eq:finalspin} as:
\begin{align}
\dfrac{1}{2}J_{\mu\nu}\Omega^{\mu\nu} &= \bm{J}\cdot\bm{\Omega}+\dfrac{1}{2}{J}_{ij}A_{CM}^{i}V_{CM}^{j} \nonumber \\
&+\dfrac{1}{2}{J}_{ij}(2V_{CM}^{i}\partial^{j}\tilde{\phi}+\partial^{i}\tilde{A}^{j})\, ,
\end{align}
which is the equation used in the main text.
%Finally, we can verify that the change of the spin tensor under the boost that we have picked is of the same order as the shift of the center of the body that accounts for 1PN the relativistic definition of the center of mass. Indeed we have:
%\begin{align}
%	\delta z_{i}={J}_{ij}V_{CM}^j/m=\epsilon_{ijk}\nu\sqrt{G_N m a(1-e^2)}\gamma^kV_{CM}^j= 2\langle r^i\delta\rangle\;,
%\end{align}

% Create the reference section using BibTeX:
\bibliography{Kozai.bib}

\end{document}